\documentclass[10pt,journal,compsoc]{IEEEtran}
\usepackage{longtable}
\usepackage{afterpage}
\usepackage[table,xcdraw,prologue]{xcolor}
\usepackage{booktabs} 
\usepackage{subfigure}
\usepackage{mathrsfs}
\usepackage[ruled,vlined]{algorithm2e}
\usepackage[utf8]{inputenc}
\usepackage{array}
\usepackage{multirow}
\usepackage{makecell}
\usepackage{amsmath,amssymb,amsfonts}
\usepackage{algorithmic}
\usepackage{graphicx}
\usepackage{textcomp}
\usepackage{tabu}
\usepackage{rotating}
\usepackage{url}
\usepackage{float}
\usepackage{caption}
\usepackage{amsthm}
\usepackage{xtab}

\begin{document}
\title{Urban Anomaly Analytics: Description, Detection, and Prediction}

\author{Mingyang~Zhang,
        Tong~Li,\IEEEmembership{Student Member,~IEEE,}
        Yue~Yu,\\
        Yong~Li,\IEEEmembership{Senior Member,~IEEE,}
        Pan~Hui,\IEEEmembership{Fellow,~IEEE}
        ~and~Yu~Zheng,\IEEEmembership{Senior Member,~IEEE,}
\IEEEcompsocitemizethanks
{\IEEEcompsocthanksitem M, Zhang, T. Li and P. Hui are with the System and Media Laboratory (SyMLab), Department of Computer Science and Engineering, Hong Kong University of Science and Technology, Hong Kong. T. Li and P. Hui are also with the Department of Computer Science, University of Helsinki, Helsinki, Finland.\protect\\
E-mail: mzhangbj@ust.hk, t.li@connect.ust.hk, panhui@cse.ust.hk

\IEEEcompsocthanksitem Y, Yu and Y. Li are with Beijing National Research Center for Information Science and Technology (BNRist), Department of Electronic Engineering, Tsinghua University, Beijing 100084, China.\protect\\
E-mail: liyong07@tsinghua.edu.cn

\IEEEcompsocthanksitem Yu Zheng is with Urban Computing Business Unit, JD Finance,
Beijing, China. \protect\\
E-mail: msyuzheng@outlook.com
}
\thanks{Manuscript received XX xx, 2019; revised xx xx, 2019.}}

%
%
%
%

\markboth{Journal of \LaTeX\ Class Files,~Vol.~14, No.~8, August~2019}%
{Shell \MakeLowercase{\textit{et al.}}: Bare Demo of IEEEtran.cls for Computer Society Journals}
%



\IEEEtitleabstractindextext{%
\begin{abstract}
Urban anomalies may result in loss of life or property if not handled properly. Automatically alerting anomalies in their early stage or even predicting anomalies before happening are of great value for populations. Recently, data-driven urban anomaly analysis frameworks have been forming, which utilize urban big data and machine learning algorithms to detect and predict urban anomalies automatically. In this survey, we make a comprehensive review of the state-of-the-art research on urban anomaly analytics. We first give an overview of four main types of urban anomalies, traffic anomaly, unexpected crowds, environment anomaly, and individual anomaly. Next, we summarize various types of urban datasets obtained from diverse devices, \emph{i.e.}, trajectory, trip records, CDRs, urban sensors, event records, environment data, social media and surveillance cameras. Subsequently, a comprehensive survey of issues on detecting and predicting techniques for urban anomalies is presented. Finally, research challenges and open problems as discussed.
\end{abstract}

\begin{IEEEkeywords}
Anomaly detection, spatiotemporal data mining, urban computing, event detection.
\end{IEEEkeywords}
}

\maketitle

\IEEEdisplaynontitleabstractindextext

%
\IEEEpeerreviewmaketitle

\section{Introduction}

\IEEEPARstart{U}{rban} anomalies are typically unusual events occurring in urban environments, such as traffic congestion and unexpected crowd gathering, which may pose tremendous threats to public safety and stability if not timely handled \cite{zhang2018detecting, zhang2019decomposition}. For example, on January 26, 2017, in Harbin, the largest city in the northeastern region of China, a single traffic incident caused a serial rear-end collision accident where eight people were killed, and thirty-two people were injured. The government then admitted that they did not detect the incident timely, which caused they could not take immediate actions to prevent the happening of the consequential tragedy. Therefore, for policymakers and government, detecting anomalies at the early stage and even predicting anomalies before happening are of the great value to prevent serious incidents from occurring. On the other hand, detecting and predicting urban anomalies are also of great importance to improve the quality of life for citizens \cite{garcia2016comparative}. For example, traffic jams are the most headache problem for most metropolises nowadays. A severe traffic jam can bring a lot of economic loss and ruin people's good moods. If most traffic jams happened in a city can be predicted or detected at the early stage, it can further be avoided becoming serious by notifying people to change their travel routes or transport. In this way, it will save people a lot of time on the commute and further improve their quality of lives.

Meanwhile, smart devices and various kinds of sensors widely located in cities collect the data produced in urban areas in real-time and form a large-scale, cross-domain and multi-view data ecosystem. These collected data, termed urban big data \cite{barkham2018urban, zheng2014urban}, differentiate from other types of data from three aspects. First, the volume of urban big data is large. Huge number of urban activities leave rich digital footprints such as trajectories of vehicles and posts on social media platforms, which compose of immense amount of urban data. Second, the forms of urban big data are various. Different sources usually produce urban data in different forms, including structured data such as human trip records and unstructured data such as surveillance videos. Moreover, urban data are associated with timestamps and location tags, which usually contain rich contextual information and bring complex temporal and spatial correlations and dependencies among different data points. Due to these unique qualities, the study of urban big data has formed an emerging research area that attracted wide interests in recent years. On one hand, new systems and algorithms have been proposed for efficient management~\cite{li2020just, li2020trajmesa, li2018efficient} and analysis~\cite{li2017mining, wang2019deep} of urban big data. On the other hand, the arising of urban big data also inspired many new applications, ranging from understanding city-scale human mobility and activity patterns~\cite{gonzalez2008understanding, zheng2008understanding, cici2015decomposition, ferreira2013visual, cranshaw2012livehoods,zhang2017regions,li2020apps}, inferring land usage and region functions~\cite{yuan2012discovering, toole2012inferring, zhang2020Multi-View} to discovering traffic problems~\cite{chawla2012inferring}, predicting air quality~\cite{zheng2013u} and diagnosing urban noise problems~\cite{zheng2014diagnosing}. 

\begin{figure}[t]
\centering
\includegraphics[width=0.45\textwidth]{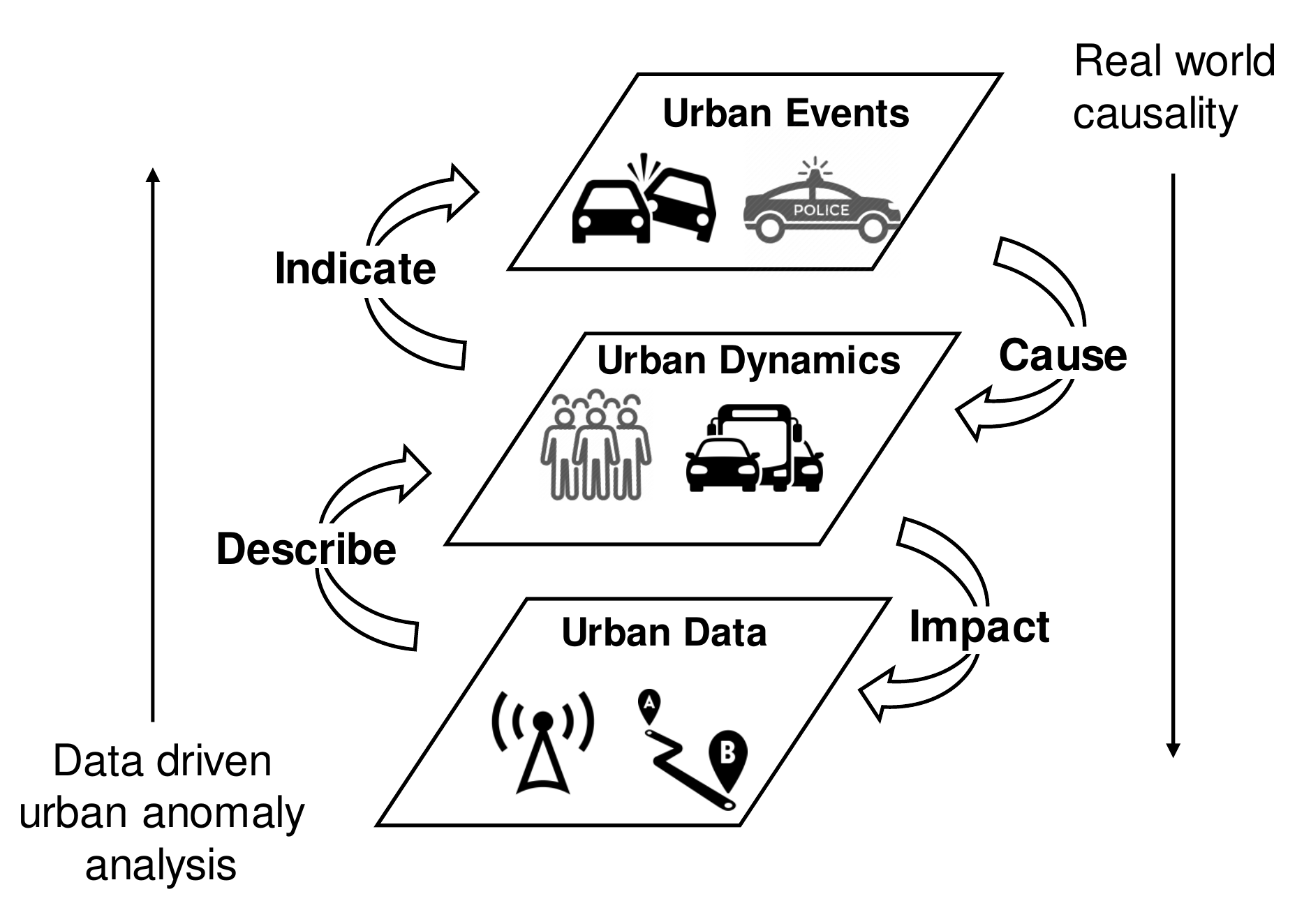}
\caption{The logic of data-driven urban anomaly analysis.}
\label{fig:logic}
\end{figure}

In this urban big data era, new urban anomaly analysis frameworks have formed as well which utilize data-driven intelligence to detect and predict urban anomalies automatically \cite{liao2010anomaly, de2016urban, cao2018voila}. Compared with traditional methods which rely on human observations and reports, the data-driven urban anomaly analysis frameworks are low-cost and more efficient \cite{zheng2014urban, kong2018lotad}. For example, many taxis are equipped with GPS devices that can report precise locations of cars in a second-level sample rate. From these location series, we can quickly know the trajectories of taxis and further infer the traffic condition in specific areas and detect traffic anomalies by looking at the speed and routing choices of cars. The logic of data-driven urban anomaly analysis is shown in Fig.~\ref{fig:logic}. We first define three concepts that establish the bridge connecting cyberspace and physical space.

\begin{itemize}

\item\textbf{Urban data} are spatiotemporal data produced by mobile devices or distributed sensors in cities. These data are usually associated with timestamps and location tags. The urban data can be in different forms and contain information of a location in a particular time or time interval.

\item\textbf{Urban dynamics} refer to the changes of urban physical status, which includes urban human mobility such as crowds or traffic flows and changes of urban environment. They are the fundamental elements of urban events and the basic description of a place's condition.

\item\textbf{Urban events} refer to social or individual activities that happen in urban areas, which are the underlying causes of urban dynamics. In the context of urban anomaly detection and prediction, we consider two types of urban events: normal urban events and abnormal urban events. The former are regular urban human activities or environment changes. The latter are unexpected irregular events, which are the targets of urban anomaly detection and prediction.

\end{itemize}

To understand the logic of data-driven urban anomaly analysis, we start with illustrating the relations among above three concepts from the view of real world causality. First, urban events are the root causes of urban dynamics. The changes of urban status, such as moving of human beings and cars, are always driven by motivations that correspond to specific urban events. For example, the traffic flows on roads around a residence area at 6pm may be caused by the event that people going home from work places. Second, urban dynamics impact urban data. For example, the gathering of crowds will make the number of cellphones connecting to the nearby cellular towers rise sharply. The logic of data-driven urban anomaly detection or prediction follows the opposite direction. First, urban dynamics are described by urban data. Urban dynamics are hard to be directly observed and understood timely and comprehensively due to their rapidly changing and spatial complexity, but they can be inferred from urban data. For example, from the loop detectors on main roads we can easily know about the traffic conditions around the city. Second, urban dynamics indicate the underlying urban events. The assumption that normal urban dynamics correspond to normal events while abnormal urban dynamics indicate urban anomalies makes the cornerstone of data-driven urban anomaly detection and prediction. By detecting abnormal patterns in urban dynamics or predict abnormal urban dynamics, we can finally achieve the goal of urban anomaly detection or prediction.

\begin{figure*}[t]
\centering
\includegraphics[width=0.9\textwidth]{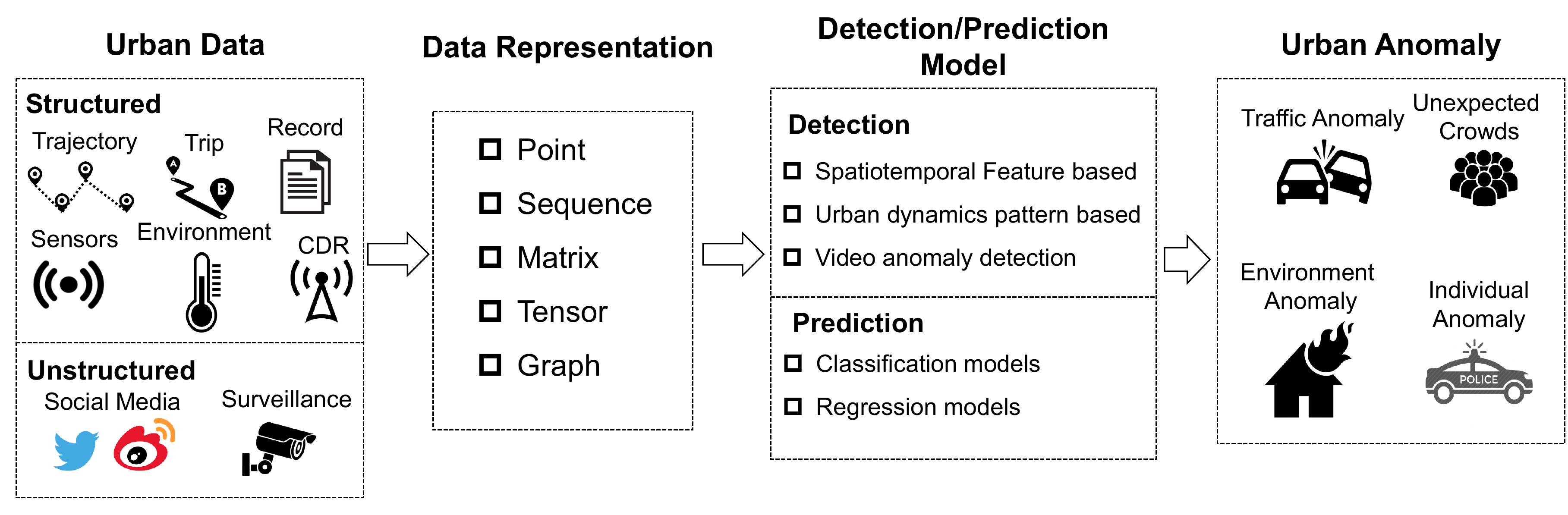}
\caption{General framework of data-driven urban anomaly analysis.}
\label{fig:framework}
\end{figure*}

Following above logic, we propose a general framework of data-driven urban anomaly analysis in Fig.~\ref{fig:framework}. Various types of urban data are first represented in specific data structures. Then the data are feed into detection or prediction models to identify or forecast the happening of concerned urban anomalies. The four components of the framework correspond to following questions we discuss in this survey. 

\begin{itemize}
\item What kinds of urban data are used and how to represent them in data structures? 
\item What are the general detection and prediction methods?
\item What kinds of anomalous events can be detected and predicted? 
\end{itemize}

To answer above questions, we systematically study relevant research in recent years and draw our conclusions. We overview the related works from three aspects, \emph{i.e.}, urban data, anomaly types and methods. First, in section~\ref{sec:data}, we describe the types of urban data commonly used and the data structures used to represent urban data. In section~\ref{sec:events}, we introduce four main types of urban anomalies studied in the existing literature, \emph{i.e.}, traffic anomaly, unexpected crowds, environment anomaly, and individual anomaly. After that, we summarize the algorithms proposed in the related literature in section~\ref{sec:alg}. In section~\ref{sec:chall}, we discuss the open challenges in urban anomaly filed. Finally, conclusions are drawn in section~\ref{sec:Conclusion}. In Table~\ref{literatures}, we list the literature in terms of what types of anomaly they detect or predict and what kinds of datasets and methods they use.

There are several surveys on related topics. Outlier detection for various types of data have been widely studied and reviewed~\cite{chandola2009anomaly,akoglu2015graph,salehi2018survey}. Our paper focuses on the scope of urban anomalous event detection, which has formed special logic and methods due to the unique attributes of urban big data and its close connection to the real-world environment. Zheng \emph{et al.}~\cite{zheng2014urban} first proposed the concept of \emph{urban computing} and discussed data-driven urban applications including urban planning, transportation, public safety and so on. This paper also involved urban anomaly detection as a child topic of urban computing, but merely reviewed the methods of traffic anomaly and disaster detection. In this survey, we greatly extend the discussion by giving a systematic formulation and review of urban anomaly from the aspects of basic logic, data types, anomaly types and methods. There are also some survey papers on traffic anomaly detection~\cite{bhowmick2018trajectory, djenouri2019survey}. Compared with these works, our survey covers a wider range of general urban anomalies and aims to provide a universal discussion on the detection and prediction logic and framework of urban anomalies. The contributions of this paper are three-fold. First, we formulated and reviewed the data-driven urban anomaly detection and prediction problem for the first time. We defined the basic concepts and proposed the general logic. Second, we systematically investigated considerable number of recent related works and proposed our taxonomies in terms of the urban data, anomaly types, detection and prediction methods. Third, we discussed the challenging problems in this area and pointed out potential research directions. 


{\tiny
\begin{table*}
\centering
   \captionsetup{font=scriptsize}
    \caption{Urban Anomaly Analysis Works. \scriptsize (Tj. is Trajectory,
    Tp. is Trip, US. is Urban Sensor, SM. is Social Media, ER. is Event Records, EV. is Environment Data. SC. is Surveillance Camera. In Urban Anomaly columns, T. is Traffic Anomaly, C. is Unexpected Crowds, E. is Environment Anomaly, I. is Individual Anomaly. In Detection columns, S. is Spatiotemoral Feature Based, P. is Urban Dynamics Pattern based. In Prediction columns, C. is Classification Methods, R. is Regression Methods.)}
 \label{literatures}
 \scriptsize
 \renewcommand{\arraystretch}{0.95}
\begin{tabular}{|c|c|c|c|c|c|c|c|c|c|c|c|c|c|c|c|c|c|c|}
\hline
\multicolumn{2}{|c|}{Literatures} & \multicolumn{8}{c|}{Urban Data} & \multicolumn{4}{c|}{Urban Anomaly} & \multicolumn{3}{c|}{Detection} & \multicolumn{2}{c|}{Prediction} \\ \hline
Name & Year & TJ. & TP. & C. & US. & ER. & EV. & SM. & SC. & T. & C. & E. & I. & S. & U. & V. & C. & R. \\ \hline
Lee et al.\cite{lee2008trajectory} &2008& $\surd$ &  &  &  &  &  &  &  &  &  &  & $\surd$ & \cellcolor[HTML]{EFEFEF}$\surd$ & \cellcolor[HTML]{EFEFEF} & \cellcolor[HTML]{EFEFEF} &  &  \\ \hline
Piciarelli et al.\cite{Piciarelli2008Trajectory} &2008&  &  &  &  &  &  &  & $\surd$ &  &  &  & $\surd$ & \cellcolor[HTML]{EFEFEF} & \cellcolor[HTML]{EFEFEF} & \cellcolor[HTML]{EFEFEF}$\surd$ &  &  \\ \hline
Li et al.\cite{li2009temporal} &2009&  & $\surd$ &  &  &  &  &  &  & $\surd$ &  &  &  & \cellcolor[HTML]{EFEFEF}$\surd$ & \cellcolor[HTML]{EFEFEF} & \cellcolor[HTML]{EFEFEF} &  &  \\ \hline
Benezeth et al.\cite{Benezeth2009Abnormal} &2009&  &  &  &  &  &  &  & $\surd$ &  &  &  & $\surd$ & \cellcolor[HTML]{EFEFEF} & \cellcolor[HTML]{EFEFEF} & \cellcolor[HTML]{EFEFEF}$\surd$ &  &  \\ \hline
Kim et al.\cite{Kim2009Observe} &2009&  &  &  &  &  &  &  & $\surd$ &  &  &  & $\surd$ & \cellcolor[HTML]{EFEFEF} & \cellcolor[HTML]{EFEFEF} & \cellcolor[HTML]{EFEFEF}$\surd$ &  &  \\ \hline
Mehran et al.\cite{Mehran2009Abnormal} &2009&  &  &  &  &  &  &  & $\surd$ &  &  &  & $\surd$ & \cellcolor[HTML]{EFEFEF} & \cellcolor[HTML]{EFEFEF} & \cellcolor[HTML]{EFEFEF}$\surd$ &  &  \\ \hline
Ge et al.\cite{ge2010top} &2010& $\surd$ &  &  &  &  &  &  &  &  &  &  & $\surd$ & \cellcolor[HTML]{EFEFEF}$\surd$ & \cellcolor[HTML]{EFEFEF} & \cellcolor[HTML]{EFEFEF} &  &  \\ \hline
Mahadevan et al.\cite{Mahadevan2010Anomaly} &2010&  &  &  &  &  &  &  & $\surd$ &  &  &  & $\surd$ & \cellcolor[HTML]{EFEFEF} & \cellcolor[HTML]{EFEFEF} & \cellcolor[HTML]{EFEFEF}$\surd$ &  &  \\ \hline
Yang et al.\cite{yang2011anomaly} &2011&  &  &  & $\surd$ &  &  &  &  &  & $\surd$ &  &  & \cellcolor[HTML]{EFEFEF} & \cellcolor[HTML]{EFEFEF}$\surd$ & \cellcolor[HTML]{EFEFEF} &  &  \\ \hline
Yang et al.\cite{yang2011anomaly2} &2011&  &  &  & $\surd$ &  &  &  &  &  & $\surd$ &  &  & \cellcolor[HTML]{EFEFEF}$\surd$ & \cellcolor[HTML]{EFEFEF} & \cellcolor[HTML]{EFEFEF} &  &  \\ \hline
Pang et al.\cite{pang2011mining} &2011& $\surd$ &  &  &  &  &  &  &  &  & $\surd$ &  &  & \cellcolor[HTML]{EFEFEF} & \cellcolor[HTML]{EFEFEF}$\surd$ & \cellcolor[HTML]{EFEFEF} &  &  \\ \hline
Ge et al.\cite{ge2011taxi} &2011& $\surd$ &  &  &  &  &  &  &  &  &  &  & $\surd$ & \cellcolor[HTML]{EFEFEF}$\surd$ & \cellcolor[HTML]{EFEFEF} & \cellcolor[HTML]{EFEFEF} &  &  \\ \hline
Zhang et al.\cite{zhang2011ibat} &2011& $\surd$ &  &  &  &  &  &  &  &  &  &  & $\surd$ & \cellcolor[HTML]{EFEFEF}$\surd$ & \cellcolor[HTML]{EFEFEF} & \cellcolor[HTML]{EFEFEF} &  &  \\ \hline
Chen et al.\cite{chen2011real} &2011& $\surd$ &  &  &  &  &  &  &  &  &  &  & $\surd$ & \cellcolor[HTML]{EFEFEF}$\surd$ & \cellcolor[HTML]{EFEFEF} & \cellcolor[HTML]{EFEFEF} &  &  \\ \hline
Saligrama et al.\cite{saligrama2012video} &2012&  &  &  &  &  &  &  & $\surd$ &  &  &  & $\surd$ & \cellcolor[HTML]{EFEFEF} & \cellcolor[HTML]{EFEFEF} & \cellcolor[HTML]{EFEFEF}$\surd$ &  &  \\ \hline
Ceapa et al.\cite{ceapa2012avoiding} &2012&  & $\surd$ &  &  &  &  &  &  & $\surd$ &  &  &  & \cellcolor[HTML]{EFEFEF} & \cellcolor[HTML]{EFEFEF}$\surd$ & \cellcolor[HTML]{EFEFEF} &  &  \\ \hline
Zhang et al.\cite{zhang2012smarter} &2012&  & $\surd$ &  &  &  &  &  &  & $\surd$ &  &  &  & \cellcolor[HTML]{EFEFEF}$\surd$ & \cellcolor[HTML]{EFEFEF} & \cellcolor[HTML]{EFEFEF} &  &  \\ \hline
Chawla et al.\cite{chawla2012inferring} &2012& $\surd$ &  &  &  &  &  &  &  & $\surd$ &  &  &  & \cellcolor[HTML]{EFEFEF}$\surd$ & \cellcolor[HTML]{EFEFEF} & \cellcolor[HTML]{EFEFEF} &  &  \\ \hline
Witayangkurn et al.\cite{witayangkurn2013anomalous} &2013&  &  & $\surd$ &  &  &  &  &  &  & $\surd$ &  &  & \cellcolor[HTML]{EFEFEF} & \cellcolor[HTML]{EFEFEF}$\surd$ & \cellcolor[HTML]{EFEFEF} &  &  \\ \hline
Pan et al.\cite{pan2013crowd} &2013& $\surd$ &  &  &  &  &  &  &  & $\surd$ &  &  &  & \cellcolor[HTML]{EFEFEF}$\surd$ & \cellcolor[HTML]{EFEFEF} & \cellcolor[HTML]{EFEFEF} &  &  \\ \hline
Li et al.\cite{Li2013Anomaly} &2013&  &  &  &  &  &  &  & $\surd$ &  &  &  & $\surd$ & \cellcolor[HTML]{EFEFEF} & \cellcolor[HTML]{EFEFEF} & \cellcolor[HTML]{EFEFEF}$\surd$ &  &  \\ \hline
Zhu et al.\cite{Zhu2013Context} &2013&  &  &  &  &  &  &  & $\surd$ &  &  &  & $\surd$ & \cellcolor[HTML]{EFEFEF} & \cellcolor[HTML]{EFEFEF} & \cellcolor[HTML]{EFEFEF}$\surd$ &  &  \\ \hline
Rozenshtein et al.\cite{rozenshtein2014event} &2014&  & $\surd$ &  &  &  &  & $\surd$ &  &  & $\surd$ &  &  & \cellcolor[HTML]{EFEFEF} & \cellcolor[HTML]{EFEFEF} $\surd$& \cellcolor[HTML]{EFEFEF} &  &  \\ \hline
Yang et al.\cite{yang2014detecting} &2014&  &  &  & $\surd$ &  &  &  &  & $\surd$ &  &  &  & \cellcolor[HTML]{EFEFEF}$\surd$ & \cellcolor[HTML]{EFEFEF} & \cellcolor[HTML]{EFEFEF} &  &  \\ \hline
Sabokrou et al.\cite{Sabokrou2015Real} &2015&  &  &  &  &  &  &  & $\surd$ &  &  &  & $\surd$ & \cellcolor[HTML]{EFEFEF} & \cellcolor[HTML]{EFEFEF} & \cellcolor[HTML]{EFEFEF}$\surd$ &  &  \\ \hline
Cheng et al.\cite{Cheng2015Video} &2015&  &  &  &  &  &  &  & $\surd$ &  &  &  & $\surd$ & \cellcolor[HTML]{EFEFEF} & \cellcolor[HTML]{EFEFEF} & \cellcolor[HTML]{EFEFEF}$\surd$ &  &  \\ \hline
Xu et al.\cite{Xu2015Learning} &2015&  &  &  &  &  &  &  & $\surd$ &  &  &  & $\surd$ & \cellcolor[HTML]{EFEFEF} & \cellcolor[HTML]{EFEFEF} & \cellcolor[HTML]{EFEFEF}$\surd$ &  &  \\ \hline
Kinoshita et al.\cite{kinoshita2015real} &2015& $\surd$ &  &  &  &  &  &  &  & $\surd$ &  &  &  & \cellcolor[HTML]{EFEFEF} & \cellcolor[HTML]{EFEFEF}$\surd$ & \cellcolor[HTML]{EFEFEF} &  &  \\ \hline
Chen et al.\cite{chen2015sensing} &2015&  & $\surd$ &  &  &  &  &  &  &  & $\surd$ &  &  & \cellcolor[HTML]{EFEFEF} & \cellcolor[HTML]{EFEFEF}$\surd$ & \cellcolor[HTML]{EFEFEF} &  &  \\ \hline
Zhang et al.\cite{zhang2015city} &2015& $\surd$ &  &  &  &  &  &  &  &  & $\surd$ &  &  & \cellcolor[HTML]{EFEFEF} & \cellcolor[HTML]{EFEFEF}$\surd$ & \cellcolor[HTML]{EFEFEF} &  &  \\ \hline
Zheng et al.\cite{zheng2015detecting} &2015&  & $\surd$ &  &  & $\surd$ &  &  &  &  & $\surd$ &  &  & \cellcolor[HTML]{EFEFEF} & \cellcolor[HTML]{EFEFEF}$\surd$ & \cellcolor[HTML]{EFEFEF} &  &  \\ \hline
Dong et al.\cite{dong2015inferring} &2015&  &  & $\surd$ &  &  &  &  &  &  & $\surd$ &  &  & \cellcolor[HTML]{EFEFEF}$\surd$ & \cellcolor[HTML]{EFEFEF} & \cellcolor[HTML]{EFEFEF} &  &  \\ \hline
Wang et al.\cite{wang2016feature} &2016& $\surd$ &  &  &  &  &  &  &  & $\surd$ &  &  &  & \cellcolor[HTML]{EFEFEF}$\surd$ & \cellcolor[HTML]{EFEFEF} & \cellcolor[HTML]{EFEFEF} &  &  \\ \hline
Wang et al.\cite{wang2016early} &2016&  &  &  & $\surd$ &  &  &  &  &  & $\surd$ &  &  & \cellcolor[HTML]{EFEFEF} & \cellcolor[HTML]{EFEFEF}$\surd$ & \cellcolor[HTML]{EFEFEF} &  &  \\ \hline
De et al.\cite{de2016urban} &2016&  &  &  & $\surd$ &  &  &  &  &  & $\surd$ &  &  & \cellcolor[HTML]{EFEFEF} & \cellcolor[HTML]{EFEFEF}$\surd$ & \cellcolor[HTML]{EFEFEF} &  &  \\ \hline
Banerjee et al.\cite{banerjee2016mantra} &2016& $\surd$ &  &  &  &  &  &  &  &  &  &  & $\surd$ & \cellcolor[HTML]{EFEFEF}$\surd$ & \cellcolor[HTML]{EFEFEF} & \cellcolor[HTML]{EFEFEF} &  &  \\ \hline
Wu et al.\cite{wu2017fast} &2017& $\surd$ &  &  &  &  &  &  &  &  &  &  & $\surd$ & \cellcolor[HTML]{EFEFEF}$\surd$ & \cellcolor[HTML]{EFEFEF} & \cellcolor[HTML]{EFEFEF} &  &  \\ \hline
Chiang et al.\cite{chiang2017btci} &2017& $\surd$ &  &  &  &  &  &  &  & $\surd$ &  &  &  & \cellcolor[HTML]{EFEFEF}$\surd$ & \cellcolor[HTML]{EFEFEF} & \cellcolor[HTML]{EFEFEF} &  &  \\ \hline
Vahedian et al.\cite{vahedian2017forecasting} &2017& $\surd$ &  &  &  &  &  &  &  &  & $\surd$ &  &  & \cellcolor[HTML]{EFEFEF} & \cellcolor[HTML]{EFEFEF} & \cellcolor[HTML]{EFEFEF} &  &  \\ \hline
Hu et al.\cite{hu2017online} &2017&  & $\surd$ &  &  &  &  & $\surd$ &  &  & $\surd$ &  &  & \cellcolor[HTML]{EFEFEF}$\surd$ & \cellcolor[HTML]{EFEFEF} & \cellcolor[HTML]{EFEFEF} &  &  \\ \hline
Zhu et al.\cite{zhu2017urban} &2017&  & $\surd$ &  &  &  &  &  &  &  & $\surd$ &  &  & \cellcolor[HTML]{EFEFEF} & \cellcolor[HTML]{EFEFEF} $\surd$ & \cellcolor[HTML]{EFEFEF} &  &  \\ \hline
Khezerlou et al.\cite{khezerlou2017traffic} &2017& $\surd$ &  &  &  &  &  &  &  &  & $\surd$ &  &  & \cellcolor[HTML]{EFEFEF} & \cellcolor[HTML]{EFEFEF}$\surd$ & \cellcolor[HTML]{EFEFEF} &  &  \\ \hline
Teng et al.\cite{teng2017anomaly} &2017&  & $\surd$ &  &  &  &  & $\surd$ &  &  & $\surd$ &  &  & \cellcolor[HTML]{EFEFEF}$\surd$ & \cellcolor[HTML]{EFEFEF} & \cellcolor[HTML]{EFEFEF} &  &  \\ \hline
Chen et al.\cite{chen2017fine} &2017&  & $\surd$ &  &  &  &  &  &  &  & $\surd$ &  &  & \cellcolor[HTML]{EFEFEF} & \cellcolor[HTML]{EFEFEF}$\surd$ & \cellcolor[HTML]{EFEFEF} &  &  \\ \hline
Tomaras et al.\cite{tomaras2017efficient} &2017&  & $\surd$ &  &  &  &  &  &  & $\surd$ &  &  &  & \cellcolor[HTML]{EFEFEF} & \cellcolor[HTML]{EFEFEF} $\surd$ & \cellcolor[HTML]{EFEFEF} &  &  \\ \hline
Xu et al.\cite{Xu2017Detecting} &2017&  &  &  &  &  &  &  & $\surd$ &  &  &  & $\surd$ & \cellcolor[HTML]{EFEFEF} & \cellcolor[HTML]{EFEFEF} & \cellcolor[HTML]{EFEFEF}$\surd$ &  &  \\ \hline
Ravanbakhsh et al.\cite{Ravanbakhsh2017Abnormal} &2017&  &  &  &  &  &  &  & $\surd$ &  &  &  & $\surd$ & \cellcolor[HTML]{EFEFEF} & \cellcolor[HTML]{EFEFEF} & \cellcolor[HTML]{EFEFEF}$\surd$ &  &  \\ \hline
Sabokrou et al.\cite{sabokrou2018deep} &2018&  &  &  &  &  &  &  & $\surd$ &  &  &  & $\surd$ & \cellcolor[HTML]{EFEFEF} & \cellcolor[HTML]{EFEFEF} & \cellcolor[HTML]{EFEFEF}$\surd$ &  &  \\ \hline
Lin et al.\cite{lin2018anomaly} &2018&  & $\surd$ &  &  &  &  &  &  &  & $\surd$ &  &  & \cellcolor[HTML]{EFEFEF} & \cellcolor[HTML]{EFEFEF} $\surd$& \cellcolor[HTML]{EFEFEF} &  &  \\ \hline
Zhang et al.\cite{zhang2018detecting} &2018&  & $\surd$ &  &  &  &  &  &  &  & $\surd$ &  &  & \cellcolor[HTML]{EFEFEF}$\surd$ & \cellcolor[HTML]{EFEFEF} & \cellcolor[HTML]{EFEFEF} &  &  \\ \hline
He et al.\cite{he2018detecting} &2018& $\surd$ &  &  &  &  &  &  &  &  &  &  & $\surd$ & \cellcolor[HTML]{EFEFEF}$\surd$ & \cellcolor[HTML]{EFEFEF} & \cellcolor[HTML]{EFEFEF} &  &  \\ \hline
Zhang et al.\cite{zhang2018deep} &2018&  &  &  &  &  &  & $\surd$ &  & $\surd$ &  &  &  & \cellcolor[HTML]{EFEFEF}$\surd$ & \cellcolor[HTML]{EFEFEF} & \cellcolor[HTML]{EFEFEF} &  &  \\ \hline
Zhu et al.\cite{zhu2018deep} &2018&  &  &  & $\surd$ & $\surd$ &  &  &  &  &  &  &  & \cellcolor[HTML]{EFEFEF}$\surd$ & \cellcolor[HTML]{EFEFEF} & \cellcolor[HTML]{EFEFEF} &  &  \\ \hline
Djenouri et al.\cite{djenouri2018outlier}&2018&&&& $\surd$ &&&&&& $\surd$ &&& \cellcolor[HTML]{EFEFEF}& \cellcolor[HTML]{EFEFEF}$\surd$ & \cellcolor[HTML]{EFEFEF} && \\ \hline
Trinh et al.\cite{trinh2019urban}&2019&&& $\surd$ &&&&&&& $\surd$ &&& \cellcolor[HTML]{EFEFEF}& \cellcolor[HTML]{EFEFEF}$\surd$  & \cellcolor[HTML]{EFEFEF} && \\ \hline
Djenouri et al.\cite{djenouri2019adapted}&2019&&&& $\surd$ &&&&&& $\surd$ &&& \cellcolor[HTML]{EFEFEF}& \cellcolor[HTML]{EFEFEF}$\surd$  & \cellcolor[HTML]{EFEFEF} && \\ \hline
Zhang et al.\cite{zhang2019decomposition}&2019&& $\surd$ &&&& $\surd$ &&&& $\surd$ &&& \cellcolor[HTML]{EFEFEF}& \cellcolor[HTML]{EFEFEF}$\surd$  & \cellcolor[HTML]{EFEFEF} && \\ \hline
Chong et al.\cite{chong2004traffic} &2004&  &  &  &  & $\surd$ &  &  &  & $\surd$ &  &  &  &  &  &  & \cellcolor[HTML]{EFEFEF}$\surd$ & \cellcolor[HTML]{EFEFEF} \\ \hline
Oh et al.\cite{oh2005real} &2005&  &  &  & $\surd$ &  &  &  &  & $\surd$ &  &  &  &  &  &  & \cellcolor[HTML]{EFEFEF}$\surd$ & \cellcolor[HTML]{EFEFEF} \\ \hline
Adbel et al.\cite{abdel2006calibrating} &2006&  &  &  & $\surd$ &  &  &  &  & $\surd$ &  &  &  &  &  &  & \cellcolor[HTML]{EFEFEF}$\surd$ & \cellcolor[HTML]{EFEFEF} \\ \hline
Ahmed et al.\cite{ahmed2012viability} &2012&  &  &  & $\surd$ &  &  &  &  & $\surd$ &  &  &  &  &  &  & \cellcolor[HTML]{EFEFEF}$\surd$ & \cellcolor[HTML]{EFEFEF} \\ \hline
Xu et al.\cite{xu2013predicting} &2013&  &  &  & $\surd$ &  &  &  &  & $\surd$ &  &  &  &  &  &  & \cellcolor[HTML]{EFEFEF}$\surd$ & \cellcolor[HTML]{EFEFEF} \\ \hline
Xu et al.\cite{xu2013genetic} &2013&  &  &  & $\surd$ &  &  &  &  & $\surd$ &  &  &  &  &  &  & \cellcolor[HTML]{EFEFEF}$\surd$ & \cellcolor[HTML]{EFEFEF} \\ \hline
Yu et al.\cite{yu2014utilizing} &2014&  &  &  & $\surd$ &  &  &  &  & $\surd$ &  &  &  &  &  &  & \cellcolor[HTML]{EFEFEF}$\surd$ & \cellcolor[HTML]{EFEFEF} \\ \hline
Copeland et al.\cite{copeland2015big} &2015&  &  &  &  &  & $\surd$ &  &  &  &  & $\surd$ &  &  &  &  & \cellcolor[HTML]{EFEFEF}$\surd$ & \cellcolor[HTML]{EFEFEF} \\ \hline
Madaio et al.\cite{madaio2015identifying} &2015&  &  &  &  &  & $\surd$ &  &  &  &  & $\surd$ &  &  &  &  & \cellcolor[HTML]{EFEFEF}$\surd$ & \cellcolor[HTML]{EFEFEF} \\ \hline
Potash et al.\cite{potash2015predictive} &2015&  &  &  &  &  & $\surd$ &  &  &  &  & $\surd$ &  &  &  &  & \cellcolor[HTML]{EFEFEF}$\surd$ & \cellcolor[HTML]{EFEFEF} \\ \hline
Chen et al.\cite{chen2016learning} &2016& $\surd$ &  &  &  & $\surd$ &  &  &  & $\surd$ &  &  &  &  &  &  & \cellcolor[HTML]{EFEFEF} & \cellcolor[HTML]{EFEFEF}$\surd$ \\ \hline
Madaio et al.\cite{madaio2016firebird} &2016&  &  &  &  &  & $\surd$ &  &  &  &  & $\surd$ &  &  &  &  & \cellcolor[HTML]{EFEFEF}$\surd$ & \cellcolor[HTML]{EFEFEF} \\ \hline
Konishi et al.\cite{konishi2016cityprophet} &2016&  &  &  &  & $\surd$ &  &  &  & $\surd$ &  &  &  &  &  &  & \cellcolor[HTML]{EFEFEF} & \cellcolor[HTML]{EFEFEF}$\surd$ \\ \hline
Huang et al.\cite{huang2016crowdsourcing} &2016&  &  &  &  & $\surd$ &  &  &  &  &  &  & $\surd$ &  &  &  & \cellcolor[HTML]{EFEFEF}$\surd$ & \cellcolor[HTML]{EFEFEF} \\ \hline
Wang et al.\cite{wang2016crime} &2016&  & $\surd$ &  &  &  &  &  &  &  &  &  & $\surd$ &  &  &  & \cellcolor[HTML]{EFEFEF} & \cellcolor[HTML]{EFEFEF}$\surd$ \\ \hline
Chojnacki et al.\cite{chojnacki2017data} &2017&  &  &  &  &  & $\surd$ &  &  &  &  & $\surd$ &  &  &  &  & \cellcolor[HTML]{EFEFEF}$\surd$ & \cellcolor[HTML]{EFEFEF} \\ \hline
Sun et al.\cite{sun2017dxnat} &2017&  &  &  & $\surd$ & $\surd$ &  &  &  & $\surd$ &  &  &  &  &  &  & \cellcolor[HTML]{EFEFEF}$\surd$ & \cellcolor[HTML]{EFEFEF} \\ \hline
Wu et al.\cite{wu2017uapd} &2017&  &  &  &  & $\surd$ &  &  &  &  &  &  & $\surd$ &  &  &  & \cellcolor[HTML]{EFEFEF} & \cellcolor[HTML]{EFEFEF}$\surd$ \\ \hline
Ren et al.\cite{ren2018deep} &2018&  &  &  &  & $\surd$ &  &  &  & $\surd$ &  &  &  &  &  &  & \cellcolor[HTML]{EFEFEF}$\surd$ & \cellcolor[HTML]{EFEFEF} \\ \hline
Singh et al.\cite{singh2018dynamic} &2018&  &  &  &  &  & $\surd$ &  &  &  &  & $\surd$ &  &  &  &  & \cellcolor[HTML]{EFEFEF}$\surd$ & \cellcolor[HTML]{EFEFEF} \\ \hline
Abernethy et al.\cite{abernethy2018activeremediation} &2018&  &  &  &  &  & $\surd$ &  &  &  &  & $\surd$ &  &  &  &  & \cellcolor[HTML]{EFEFEF} & \cellcolor[HTML]{EFEFEF}$\surd$ \\ \hline
Kumar et al.\cite{kumar2018using} &2018&  &  &  &  &  & $\surd$ &  &  &  &  & $\surd$ &  &  &  &  & \cellcolor[HTML]{EFEFEF}$\surd$ & \cellcolor[HTML]{EFEFEF} \\ \hline
Yuan et al.\cite{yuan2018hetero} &2018&  &  &  & $\surd$ &  &  &  &  & $\surd$ &  &  &  &  &  &  & \cellcolor[HTML]{EFEFEF} & \cellcolor[HTML]{EFEFEF}$\surd$ \\ \hline
Huang et al.\cite{huang2018deepcrime} &2018&  &  &  &  & $\surd$ &  &  &  &  &  &  & $\surd$ &  &  &  & \cellcolor[HTML]{EFEFEF}$\surd$ & \cellcolor[HTML]{EFEFEF} \\ \hline
Huang et al.\cite{huang2019mist} &2019&  &  &  &  & $\surd$ &  &  &  &  &  &  & $\surd$ &  &  &  & \cellcolor[HTML]{EFEFEF}$\surd$ & \cellcolor[HTML]{EFEFEF} \\ \hline

\end{tabular}
\end{table*}
}
\section{Urban data}

\label{sec:data}
\subsection{Tyeps of urban data}
\label{sec:datatype}
The datasets are the core part of data-driven urban anomaly detection and prediction. Nowadays, various data can be collected from electronic devices. One of the most important sources is smartphones. For example, when users make phone calls or access cellular networks, their locations will be recorded and can be used to track user mobility and even estimate the density of populations \cite{dong2015inferring}. The text and pictures posted on social media by smartphones are also ideal semantic descriptions of the urban environment, which can help to understand the events happening around. Also, electronic sensors distributed around urban areas are the other important sources of urban data. For example, the surveillance cameras can collect surrounding scenes and traffic sensors provide real-time traffic status. In this section, we classify urban datasets into eight categories in terms of data types and attributes. six of them are structured data, \emph{i.e.}, trajectory, trip records, urban sensor records, and mobile phone call detail records (CDRs), event records and environment data. The other two are unstructured data, \emph{i.e.}, social media data and surveillance camera data.

\subsubsection{Trajectory}
A trajectory consists of a series of time-location records, which are reported by GPS devices in a sample rate of the second level. The trajectory datasets provide the most detailed and comprehensive records of object movements. One of the primary sources of trajectory data is the taxis with GPS equipment. As one of the most critical urban transport, taxis are widespread in urban areas and run for almost 24 hours every day. Lots of works are based on taxi trajectories\cite{chen2011real, castro2013taxi, zhang2015city, zhu2017urban, li2019driving}. In practice, trajectory datasets can be used to detect traffic anomalies, unexpected crowds, and even individual anomalies. For example, gathering events are detected by predicting human mobility using trajectory data~\cite{vahedian2017forecasting}, traffic incident is caught based on taxi trajectories~\cite{kinoshita2015real} and taxi driving fraud is identified by detecting trajectory outliers~\cite{ge2011taxi}.

We summarize popular trajectory datasets used in existing literature in Table~\ref{trajectory}. These datasets are collected from metropolitan areas, especially the areas of a large population such as Beijing, Shanghai, and Singapore. The duration of datasets varies from days to years. The sample rate is usually at the second or minute level, which is considered dense enough to track the mobility of vehicles. The quantity of datasets is given by the number of trajectories(T) or the location points(P) reported by GPS.

\begin{table*}[ht]
\centering
    \caption{Trajectory datasets}
    \small
    \label{trajectory}
    \begin{tabular}{ccccc}
    \toprule
    Dataset & Duration& Sample Rate (s/point) & \#T/\#P ($\times 10^3$) & \#Vehicles \\ \hline
    Porto\cite{wu2017fast}   & 1.5 years   & 15    & 486 T                & 442              \\ \hline
    Shanghai$^1$\cite{wu2017fast}     & 10 days    & 10     & 757 T      & 13650            \\ \hline
    Beijing$^1$\cite{lin2018anomaly}  & 1 week    & -       & 450 P         & 8940             \\ \hline
    Singapore\cite{chiang2017btci}& 2 months      & -    & 50 P            & -                \\ \hline
    Shanghai$^2$\cite{zhang2015city}  & 2 years    & - & 10,000,000 P      & 10000            \\ \hline
    Hanzhou\cite{zhang2015city}      & 1 year       & -       & 3,000,000 P          & 5000             \\ \hline
    Beijing$^2$\cite{pan2013crowd}   & 2 months    & 70.5     & 19,455 T            & 13597            \\ 
    \bottomrule
    \end{tabular}
\end{table*}

\subsubsection{Trip records}

Trip records are usually collected from taxis and sharing bike systems. Each trip record contains the start location, end location, trip distance, and trip duration. One of the primary applications of vehicle trip data is to understand human mobility in urban areas. For example, the total number of taxi trips from one region to another region reflects the number of people moving from one area to another to some extent. Hence, people gathering events can be discovered by monitoring taxi trips.

\begin{table}[h]
\centering
\caption{Trip Record Datasets}
\small
\label{trip}
\begin{tabular}{cccc}
\toprule
Dataset     & Duration & \#Trips ($\times 10^6$) & \#Vehicles \\ \hline
NYC-taxi$^1$~\cite{lin2018anomaly}      & 1 year  & 3          & -                   \\ \hline
Washington~\cite{ceapa2012avoiding} & 3 year      & 8                 & 3296                \\ \hline
NYC-bike~\cite{zheng2015detecting}  & 1 year     & 8                     & 6811                \\ \hline
NYC-taxi$^2$~\cite{zheng2015detecting}  & 1 year    & 165    & 14144               \\ \hline
San Francisco~\cite{li2009temporal} & 30 days      & 0.8        & 500                 \\ 
\bottomrule
\end{tabular}

\end{table}

The most commonly used trip record datasets are listed in Table~\ref{trip}. These datasets are published by city taxi or public sharing bike operators and updated every month or season. The duration of these datasets is up to years, except the bike trip dataset from San Francisco. In taxi trajectory datasets, the start and end locations are usually given as the street name or block name, while
in sharing bike systems the locations are bike stations. Each dataset contains millions of trip records produced by thousands of unique vehicles.

\subsubsection{CDRs}

Mobile phone call detail records (CDRs) include the time and location information of phone calls~\cite{blondel2015survey}. CDR data use the positions of the associated base stations as user locations. Besides, the time interval between two phone calls made by one user is usually up to hours or even days, which makes CDRs sparser than trip data. However, since the penetration of mobile phones and the vast number of phone calls made every day, CDRs are more accessible and of a large volume. With these advantages, CDRs are widely used to estimate human mobility and population distributions \cite{ranjan2012call, zhang2014exploring, trestian2009measuring}. The typical CDR datasets are listed in Table~\ref{cdr}. These datasets are collected by the internet service providers (ISPs) and contain millions of users with the duration from months to one year.

\begin{table*}[ht]
\centering
\caption{CDR Datasets}
\small
\label{cdr}
\begin{tabular}{cccc}
\toprule
Dataset  & Duration & \# of users or divices($\times 10^6$) & \# of records ($\times 10^6$) \\ \hline
San Francisco\cite{ranjan2012call} & 1 month  & 1                   & -             \\ \hline
Massachusetts\cite{hoteit2014estimating} & 4 months  & 1                   & -             \\ \hline
Shen Zhen\cite{zhang2014exploring} & 1 year  & 10                   & 435             \\ \hline
Haiti\cite{bengtsson2011improved} & 6 weeks  & 1.9                & -             \\ \hline
Senegal\cite{dong2015inferring} & 5 months  & 0.05               & -             \\ \hline
Los Angles\cite{isaacman2011ranges} & 140 days  & 0.2               & 321            \\ \hline
New York\cite{isaacman2011ranges} & 140 days  & 0.15               & 223            \\ 
\bottomrule
\end{tabular}
\end{table*}

\subsubsection{Urban sensing data}
Apart from mobile phones and GPS devices, there are also many sensors distributed around urban areas to collect urban data. The most common sensors are the loop detectors on roads, which are installed underneath pavements at around half a mile intervals. Loop detectors record the vehicles passing that location. The records can be utilized to evaluate vehicle speeds and road conditions \cite{xu2013genetic, xu2013predicting, oh2005real}. Besides, public transport card reading machines in bus and subway stations are another important user sensors that record the volume of people flows \cite{wang2016early}. These various sensors can reflect the urban dynamics of one location from different perspectives.

\subsubsection{Event records}
For specific types of urban anomalies, there are some datasets comprehensively recording the event time, locations and descriptions. 
For example, the traffic accident records provided by traffic management departments are available in several cities~\cite{chong2004traffic,zhu2018deep,chen2016learning}. These datasets are usually used for detection or prediction result validation.
Crime records provided by police offices are also used in some works~\cite{wang2016crime,huang2018deepcrime,huang2019mist}, which usually include the time, location, category and severity information. As a special kind of urban anomaly, crime events have close relation with location properties such as average income and population density. Therefore, the crime records are commonly adopted for neighborhood crime rate prediction~\cite{wang2016crime,huang2018deepcrime}.

\subsubsection{Environment data}
Concerning environment anomaly detection in urban areas, environment data are the primary data source. The environment information such as building conditions can be used to evaluate fire risk in urban areas \cite{singh2018dynamic, madaio2016firebird, madaio2015identifying}. Also, in the case of water system monitoring, the test results of water samples are used~\cite{chojnacki2017data, potash2015predictive, kumar2018using}.  On the other hand, the happenings of other kinds of anomalous events are sometimes affected by environmental factors as well. Thus, environment data, as the extra information, are usually used to assist anomalies detection and prediction. For example, the weather data are widely used in traffic anomaly detecting since the bad weather is a primary cause of traffic accidents~\cite{xu2013genetic, abdel2006calibrating, yu2014utilizing, chong2004traffic}.

\subsubsection{Social media}
Social media data from Twitter and Weibo are widely employed for event discovering as well~\cite{lin2016exploring}. However, for urban anomalous events, social media datasets are usually not used separately. For example, the topic distribution of social media data is used as a feature and together with other types of urban data for multi-view anomaly detection and prediction~\cite{zheng2015detecting, teng2017anomaly}. In some literature, social media data are also used to get a semantic understanding of detected events~\cite{pan2013crowd}.

\subsubsection{Surveillance Camera}

 Surveillance camera plays a vital role in capturing and monitoring human mobility. Detecting abnormal behaviors such as traffic peccancy from surveillance camera videos is an important research area that attracted wide interests. We list widely used camera surveillance datasets in Table~\ref{camera}. The subway dataset consists of two videos,  `entrance gate' (1 hour 36 minutes long with 144,249 frames) and `exit gate' (43 minutes long with 64,900 frames) \cite{Adam2008Robust}. The UCSD dataset collects the videos from walkways in the campus of University of California, San Diego. The data are split into two subsets, Peds1 and Peds2. Peds1 records the people walking towards and away from the camera and contains 34 training video samples and 36 testing video samples. Peds2 records the pedestrians parallel to the camera plane and contains 16 training video samples and 12 testing video samples~\cite{ucsd}. The VIRAT dataset~\cite{Oh2011A} consists of stationary ground camera data over 25 hours and 16 different scenes of high resolution. The CUHK dataset~\cite{Lu2014Abnormal} contains 15 sequences of 2 minutes for each. In this dataset, there are 14 types of abnormal events including running, throwing objects, loitering and so on. 

\begin{table}[ht]
\centering
\caption{Camera Records Datasets}
\small
\label{camera}
\begin{tabular}{ccc}
\toprule
Dataset & \# of Frames  & Resolution \\ \hline
Subway-entrance\cite{Adam2008Robust}      & 144,249      & $512\times 384$                    \\ \hline
Subway-exit\cite{Adam2008Robust} &  64,900           & $512\times 384$                    \\ \hline
UCSD-Ped1\cite{ucsd}  & 14,000            & $158\times238$                \\ \hline
UCSD-Ped2 \cite{ucsd}  & 5,600            & $320\times 240$                 \\ \hline
VIRAT\cite{Oh2011A} & 37,500-45,000           & $1920\times 1080$                               \\ \hline
CUHK \cite{Lu2014Abnormal} &  35,240           & -                       \\ 
\bottomrule
\end{tabular}
\end{table}

\subsection{Urban data representation}

\begin{figure*}[t]
\centering
\includegraphics[width=0.7\textwidth]{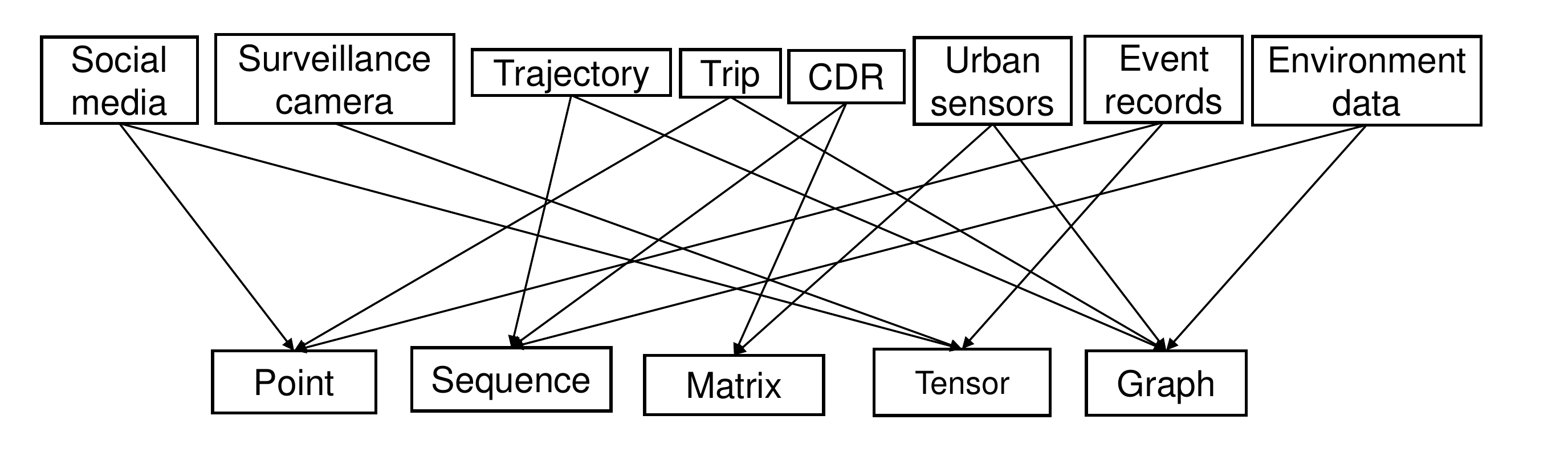}
\caption{Data structures to represent urban data.}
\label{fig:repr}
\end{figure*}

For urban data to be processed by detection and prediction models, they must be represented by data structures. The choices of data structures depend on the natural properties of urban data. One common property of above mentioned six types of urban data is that they are  associated with timestamps and locations. This kind of data are generally termed as spatiotemporal data, which differ from other data types because of the presence of correlations in both spatial and temporal domains~\cite{atluri2018spatio}. Therefore, the representation of urban data need to take their spatial relations and temporal relations into consideration. In this subsection, we discuss five data structures that are commonly used to represent urban data, which are respectively point, sequence, matrix, tensor and graph. We will discuss the characteristics of each data structure and what types of urban data they are used to represent as shown in Fig.~\ref{fig:repr}.
\begin{itemize}

\item The simplest way is to represent urban data instances with independent points. In this way, neither of the spatial and temporal relations between urban data instances is considered. Points are usually used to represent passively collected urban data, such as social media and taxi trips. For those types of urban data, The time intervals between consecutive data instances are not fixed and in the same time, the locations where data instances are produced are also uncertain, which make it hard to model their temporal dependency and spatial relations.

\item Some urban data are generated in a specific sample rate, which naturally form time series that can be represented by sequences. For example, a trajectory contains a sequence of GPS locations that generated in every few seconds.  Some other urban data, such as trips and CDRs, are a set of independent event records, which can also be modeled as sequences by counting the number of events that fall into consecutive time intervals. Comparing with points, sequences reserve the temporal orders of urban data instances but also lose the spatial relations.

\item If the urban data are multivariate time series, they are usually truncated into fixed length in time dimension and represented as matrices. The second dimension of the matrix can be regions~\cite{chen2017fine} or features~\cite{yang2014detecting}. In the former case, even multiple locations are considered in this representation, the relations between locations are still neglected. Therefore, a matrix representation still can only reserve temporal dependencies.

\item Urban data usually contain multidimensional information including times, locations and features, which can represented by tensors.
In practice, 3D tensors are most commonly used. For example, in some works~\cite{zhang2017deep} urban areas are divided into $n\times n$ grids and sequential observed data in these grids form a tensor. In this case, the geographic adjacency relations are reserved by the representation. In some other works~\cite{lin2018anomaly,chen2017fine}, the three dimensions are respectively time, location and multivariate features such as observations from different sensors.

\item Above data structures can reserve very limited spatial relations between urban data instances. The graph is utilized in some works to overcome this shortcoming~\cite{rozenshtein2014event,de2016urban,teng2017anomaly}. For example, trip data can be represented with graphs that have locations as nodes, and the weights of edges are decided by traffic flow between regions. And data collected from distributed urban sensors can also be represented as graphs with locations as nodes and geographic relations as edges.
\end{itemize}
Different data representations usually lead to different detection or prediction models. For example, by representing urban data as sequences, urban anomaly detection can be transformed to classic time series outlier detection problem~\cite{ranshous2015anomaly}. With a matrix or tensor representation, recent deep learning techniques such as Convolutional Neural Network(CNN) can be applied to process urban data. In section \ref{sec:alg}, we will discuss the detection and prediction models in details.

\section{Urban Anomalous Events}
\label{sec:events}

The events happening in urban areas can be simply classified into two types, normal events, and abnormal events. The normal events refer to regular activities following certain patterns or rules, while the abnormal events refer to incidental events happening by accident. For example, crowds gathering at rush hours around a subway station is a normal event since it follows a regular pattern occurring periodically every weekday. A pop concert held around can also cause a significant rise in subway traffic, but it should be considered as an anomalous event since it rarely happens and follows an irregular pattern. In this section, we will discuss four kinds of urban anomalies that are most commonly studied, \emph{i.e.}, traffic anomaly, unexpected crowds, environment anomaly, and individual anomaly.

\subsection{Traffic Anomaly}
The traffic is of vital importance for the daily lives of citizens. Detecting and predicting traffic anomalies attract a lot of researchers \cite{oh2005real, abdel2006calibrating, liu2011discovering, Xu2018Dual}. 
Traffic anomalies mainly have two types. The first one is traffic congestion that is usually caused by traffic incidents or traffic overload. Traffic congestion will cause the slowdown in traffic speed or increase in traffic volume on specific roads, but these effects only last for a short time like a few minutes or hours. The other one is road management, like maintenance or closure of roads, which usually causes a sharp drop in traffic volume and the effect will keep for a longer time.

The works on traffic anomaly analysis can be classified into two types: local traffic anomaly analysis and group traffic anomaly analysis. The first type of works treat the road network as a combination of independent road segments and detect or predict the anomalies for each road segment. In these works, traffic features such as vehicle speed are first extracted for different road segments. Then general anomaly detection methods~\cite{chandola2009anomaly} such as statistical methods~\cite{dong2015inferring, yang2014detecting, wang2016feature} and neighborhood based methods~\cite{djenouri2019adapted} are applied in the feature space to identify traffic anomalies. The second type of works consider the road segments are not independent and traffic anomalies usually affect multiple road segments instead of merely one road segment. In other words, if the road network is regarded as a graph, a traffic anomaly is more likely an anomalous subgraph instead of an abnormal edge. Based on this consideration, some works detected a group of connected road segments as an abnormal group\cite{pan2013crowd,chiang2017btci}. In~\cite{chawla2012inferring,liu2011discovering}, the authors further explored the causal interactions among road segments and identified the root cause of traffic anomalies.

\subsection{Unexpected Crowds}

Unexpected crowds in urban areas are one of the major threats to public safety. For example, on Dec. 31th, 2014, more than 300,000 people flowed into the Bund in Shanghai for the light show on New Year’s Eve. The volume of crowds highly exceeds the expectation of organizers and the overcrowding led to a tragic stampede and caused 36 people killed and 49 injured in the end. Such accidents could be prevented if the gathering of people is detected or predicted in its early stage \cite{pan2013crowd, dong2015inferring}. On the other hand, the increase in people density in a region usually can be reflected in the urban data domain, such as the sudden increase in the number of cellular subscribers for the base stations around that area and the unexpected rise in the number of exiting passengers of nearby subway stations~\cite{ceapa2012avoiding}. These abnormal changes in urban data can help to discover the happening of unexpected crowds.

One significant difficulty to detect or predict unexpected crowds is the limited number of recorded events, which makes it hard to evaluate the effectiveness of different detecting and predicting algorithms. In practice, existing works usually use festival celebration, pop concerts, and sports matches as abnormal events, which can be discriminated by their effect scope. For example, concerts and matches only have a local influence, while the other events like festivals and extreme weather usually cause unusual changes of urban dynamics in city-scale.

\subsection{Environment Anomaly}
Urban environment anomalies are also a significant category of anomalous urban events and highly related to public safety. For example, fire incidents in cities are a severe threat to people's lives and property \cite{madaio2016firebird, singh2018dynamic}. Besides, the pollution of the water system is another kind of urban environment anomaly that may harm residents' health \cite{chojnacki2017data, kumar2018using}. 

Unlike other kinds of urban anomalies, environment anomalies are mainly caused by environment changes instead of large-scale human activities and usually do not show significant signs before happening. Hence, instead of directly detecting and alerting the urban anomalous, current works of environment anomalies focus on evaluating the risk or tracing the causes. For example, Micheal et al. \cite{madaio2016firebird} evaluated whether a building has a risk of fire based on the building condition information. Alex et al. \cite{chojnacki2017data} explored the residential water contamination based on water sample tests. 

\subsection{Individual Anomaly}
Traffic anomalies and unexpected crowds usually have a lot of participators and have a relatively large-scale impact. However, there are still some urban anomalies that are caused by abnormal individual activities and have less public influence. For example, and some criminal or illegal activities such as taxi fraud~\cite{ge2011taxi} and illegal parking~\cite{he2018detecting}.

In this survey, we term such events as individual anomalies and also consider it a critical anomaly category. There are two reasons.
On the one hand, the detection of such anomalies can help to discover criminal or illegal activities and protect urban security. On the other hand, a large-scale anomalous event is consist of a lot of individual anomalies. Analyzing the spatial and temporal interactions of a lot of individual anomalies can also help to detect other kinds of anomalies.

\section{Algorithms}
\begin{figure}[t]
\centering
\includegraphics[width=0.46\textwidth]{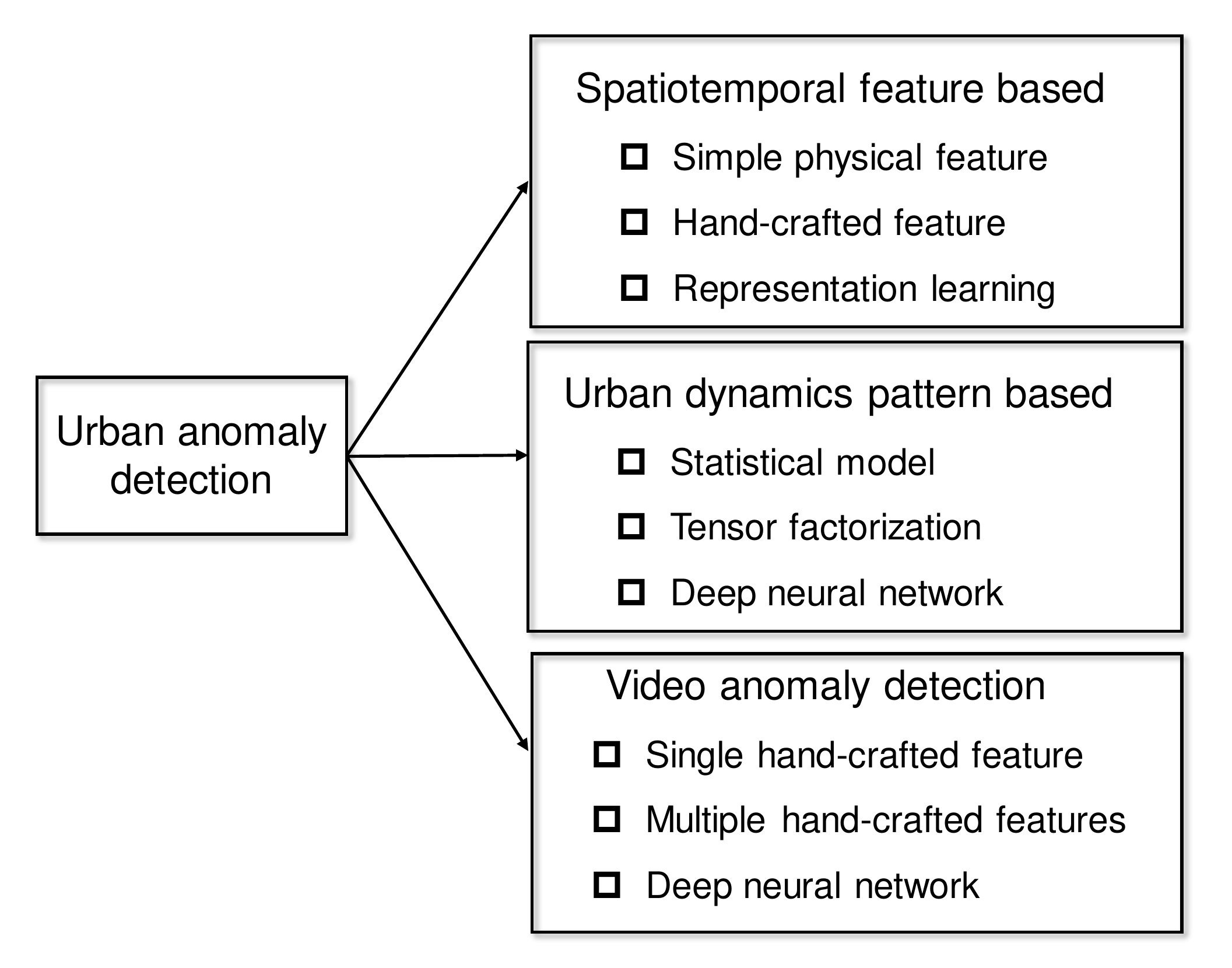}
\caption{Classification of urban anomaly detection algorithms.}
\label{methodclassification}
\end{figure}

\label{sec:alg}
\subsection{Anomaly Detection}

In this section, we will discuss the state-of-art algorithms on urban anomaly detection. While urban anomaly detection lies in the scope of outlier detection which has been widely studied and reviewed~\cite{han2011data,chandola2009anomaly}, it has particularities because of the complexity of urban data and close connection with physical urban environment. We classify the algorithms into three main groups including spatiotemporal feature based,  urban dynamic pattern based and video anomaly detection methods. An overview of our classification is shown in Fig.~\ref{methodclassification}. 

\subsubsection{Spatiotemporal Feature Based}

\begin{figure*}[t]
\centering
\includegraphics[width=0.8\textwidth]{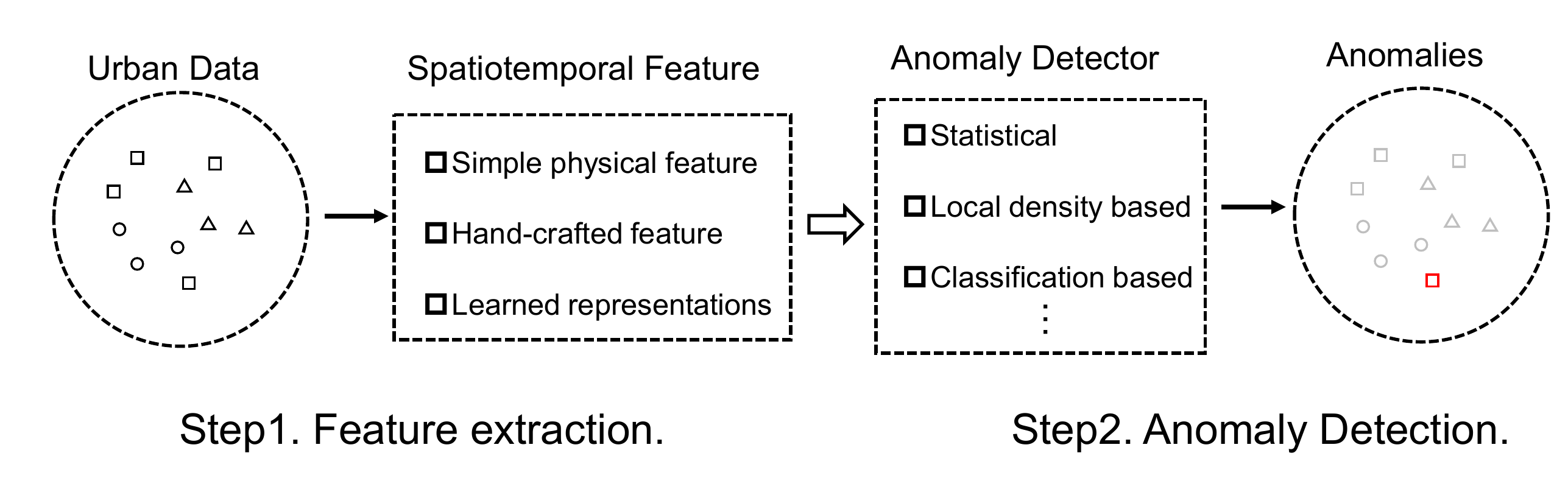}
\caption{Spatiotemporal feature based urban anoamly detection.}
\label{fig:stf}
\end{figure*}

Spatiotemporal feature based methods reduce the urban anomaly detection problem to a classical anomaly detection problem. Classical anomaly detection algorithms can not be directly applied to urban anomaly detection problem because urban data are affected by complex urban environmental factors and contain enormous amount of redundant information. To cope with this problem, essential information needs to be extracted from urban data to construct or learn a feature to embed the data instances into a general feature space. The general process of spatiotemporal feature based methods is shown in Fig.~\ref{fig:stf}, which consists of two cascaded steps. In the first step, spatiotemporal features are extracted from urban data. In second step, the features are feed into an outlier detector to identify anomalies. 

There are three levels of spatiotemporal features adopted by existing works. The first level is simple physical features, which are direct observations from urban data and usually have explicit physical meanings such as vehicle speed and travel time. The second level is hand-crafted features that constructed from urban data based on specific definitions, such as spatiotemporal similarity and driving routing pattern. And the last level is features learned from urban data by representation learning methods such as subspace learning and manifold learning methods.

\paragraph{Simple physical feature} Simple physical features are widely used because they are easy to access and have clear real world meanings. In~\cite{wang2016feature, chiang2017btci}, the average vehicle speed is used to detect traffic accidents or congestion. They divided urban road networks into small segments and estimated the traffic flow speed based on trajectories in a small time slot. In~\cite{wang2016feature}, Wang \emph{et al.} computed the change rate of traffic flow speed and regarded an anomaly occurring when the change rate exceeds a threshold. In~\cite{chiang2017btci}, Chiang \emph{et al.} derived a congestion score based on vehicle speed to detect traffic congestion. Additionally, some existing works use travel distance as a feature to detect urban anomalies. As for the trips of similar start point and destination, their travel distance should also be similar and around a constant. Based on this fact, Zhang \emph{et al.}~\cite{zhang2012smarter} proposed a framework to detect taxi trips with abnormal travel distance and further inferred traffic congestion based on abnormal trips. Similarly, in~\cite{ge2011taxi}, the travel distance was used as the evidence for taxi fraud detection.  As for trajectory outlier detection, the moving direction is another useful feature. In~\cite{ge2010top}, Ge \emph{et al.} proposed a system calculating an outlier score for a trajectory based on its moving direction sequence.

\begin{figure}[t]
\centering
\includegraphics[width=0.5\textwidth]{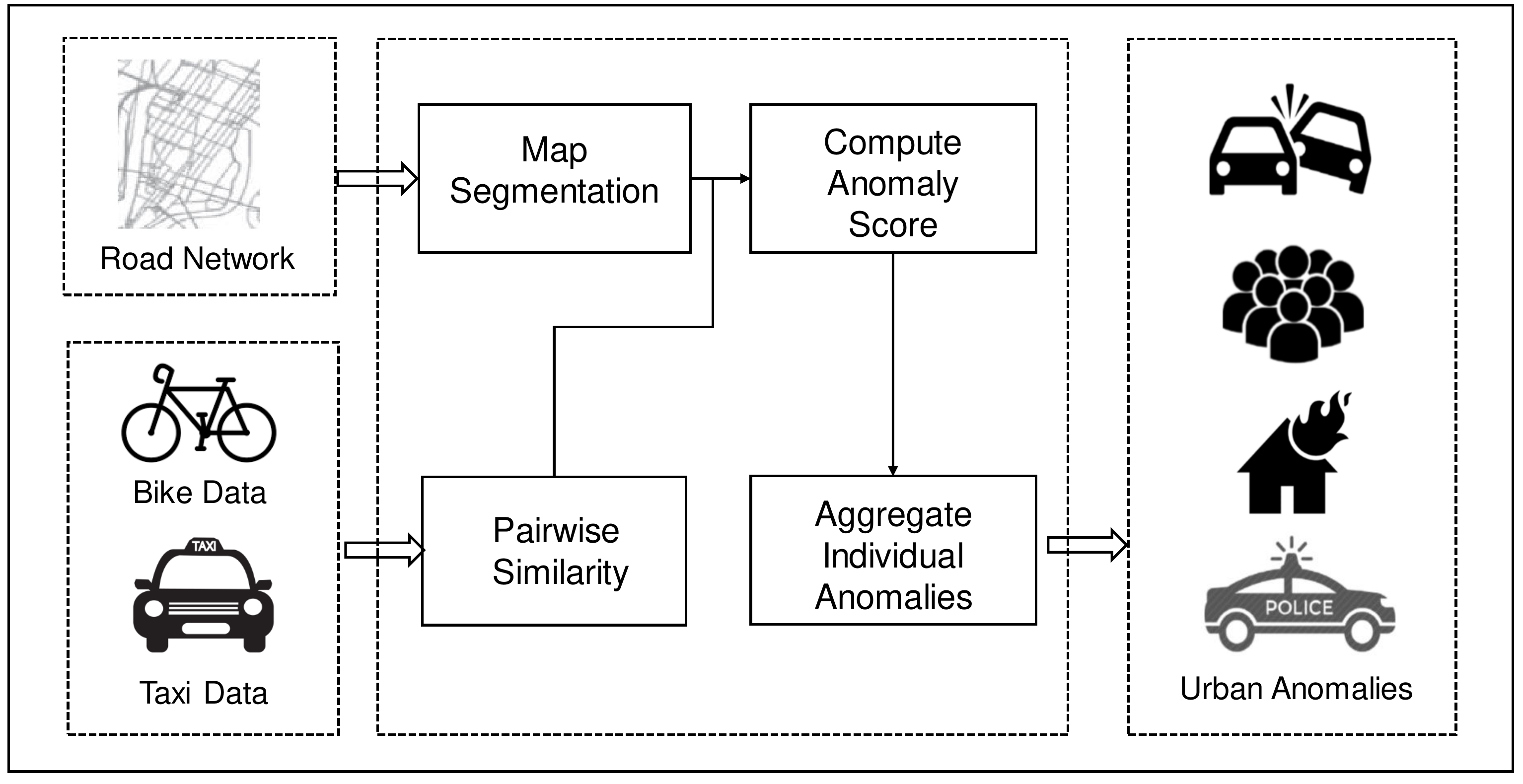}
\caption{The framework of the spatiotemporal similarity based method proposed by Zhang \emph{et al.}~\cite{zhang2018detecting}.}
\label{fig:fbexample}
\end{figure}

\paragraph{Hand-crafted feature}
Higher-level human-designed features are also used for urban anomaly detection. A widely considered feature is temporal or spatial similarity. An example~\cite{zhang2018detecting} is shown in Fig.~\ref{fig:fbexample}. This work considers the spatiotemporal similarity among urban regions as the features. First, urban area is partitioned into small regions based on road network. Then for each region the authors calculate a similarity score between this region and all other regions from their historical taxi and bike trip records. Li \emph{et al.}~\cite{li2009temporal} represented road networks as a directed graph and computed the similarity between edges. They considered every edge should have stable neighbors in the feature space and detected traffic anomalies by monitoring the change of edge neighbors. In~\cite{dong2015inferring}, Dong \emph{et al.} considered temporal similarity. They first identified anomalous users by comparing the similarity between an individual's current trajectory and his/her historical trajectories. They then detected unexpected crowds by searching gathered abnormal individuals. In the case of traffic anomaly detection, routing behavior can also be modeled as a feature. Pan \emph{et al.}~\cite{pan2013crowd} defined the routing pattern between two locations as a vector of the traffic volume on each path that connects the two locations. In~\cite{ge2011taxi}, the authors represented a trajectory as a sequence of symbols of the passed locations. They then assigned a codeword to each symbol. By this meaning, the anomalous trajectories have unusually high code cost based on coding theory. In~\cite{djenouri2018outlier,djenouri2019adapted}, the authors proposed flow distribution probability based on the traffic flow during a time interval at a location, then traffic anomalies are detected by applying general outlier detection methods on the database of flow distribution probabilities.

\paragraph{Representation learning} While simple physical features and hand-crafted features are easy to access and understood, they can hardly reserve the complex spatial and temporal correlations of urban data instances. In order to obtain comprehensive spatiotemporal features from urban data, representation learning methods are adopted in many works. 

Principal Component Analysis (PCA) and its variants are the most popular subspace learning methods. The principal idea of PCA is to use an orthogonal transformation to convert linearly correlated data into linearly uncorrelated data. A variant of PCA is Robust PCA~\cite{candes2011robust} that can tolerate non-Gaussian noises. In~\cite{chawla2012inferring}, PCA was applied on a link-time matrix which shows the traffic volume on different roads in time windows to detect anomalous routes. Then the root cause of traffic anomaly can be located by exploring the linkage relations between roads. Yang \emph{et al.}~\cite{yang2014detecting} introduced the Bayesian Robust PCA~\cite{ding2011bayesian} algorithm to co-factorize multiple traffic data streams to learn shared features of different traffic measurement data. A commonly used manifold learning algorithm is Locally Linear Embedding (LLE)~\cite{roweis2000nonlinear}. The LLE algorithm learns the representation of a data point by constructing it as the linear combination of its $K$ nearest neighbors. The weights are derived in the sense of least square construction error, and the vector of weights is regarded as features of the original data point. Yang \emph{et al.}~\cite{yang2011anomaly2} represents the traffic flow collected from distributed sensors in a time window as a matrix. They then apply the LLE and PCA algorithm together to obtain an LLE-PCA feature~\cite{yang2007manifold} for abnormal event detection. 

\begin{figure}[t]
\centering
\includegraphics[width=0.48\textwidth]{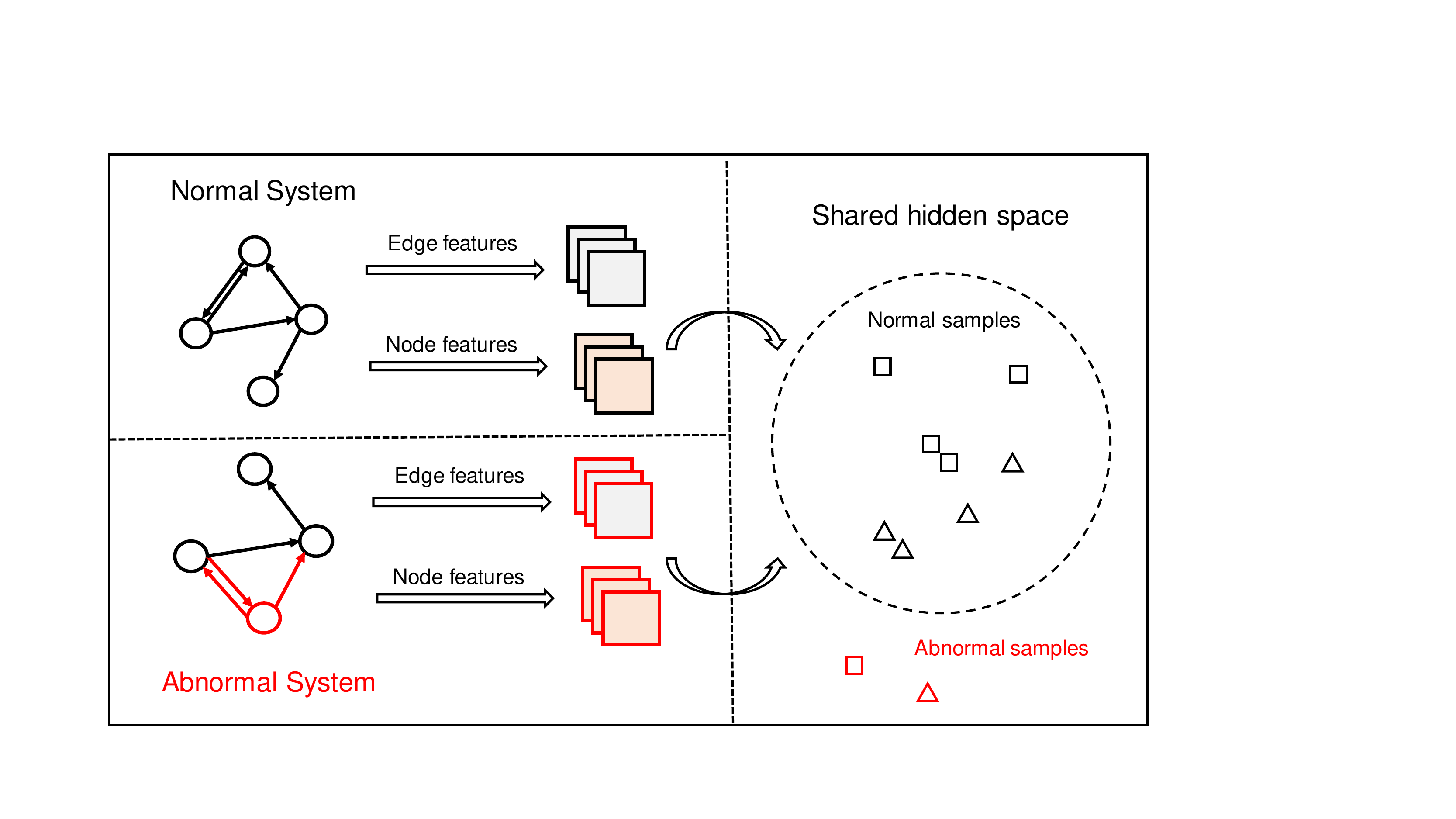}
\caption{The framework of multi-view learning method proposed by Teng \emph{et al.}~\cite{teng2017anomaly}.}
\label{fig:gbexample}
\end{figure}

Multi-view learning methods are adopted for learning spatiotemporal features across multiple datasets in~\cite{teng2017anomaly}. As shown in Fig.~\ref{fig:gbexample}, a dynamic network is first constructed based on urban data. The dynamic network contains a series of directed weighted graphs which represent urban dynamics in consecutive time slots. The nodes in graphs stand for geographic regions, and the edges represent the traffic flows between different regions with the weight set as the traffic volume. The spatial feature such as the twitter topic distribution of each region is assigned as node properties. Then a multi-view hypersphere learning algorithm was proposed to learn a latent representation of each node that fuses both node and edge side information, and anomalous nodes were detected in the latent space.

After obtaining appropriate features, urban anomaly detection can be transformed into classical anomaly detection problems. Classical anomaly detection methods can be grouped into several categories, statistical methods, classification based methods, nearest neighborhood-based methods, clustering based methods\cite{chandola2009anomaly}. The first two categories' methods are most frequently used in the second step of spatiotemporal feature based urban anomaly detection methods. The principle of statistical anomaly detection methods is to estimate the distribution of data and consider the instances of low probabilities as anomalies. There are two common methods to estimate the data distribution, parametric methods and nonparametric methods. The gaussian-based parametric method is a main method used in urban anomaly detection~\cite{wang2016feature,zhang2012smarter,banerjee2016mantra,ge2010top,dong2015inferring,pan2013crowd}. The Gaussian-based model assumes data follow a Gaussian distribution and estimates the parameters with Maximum Likelihood Estimates (MLE). Based on the estimated parameters, the anomaly score of an instance can be defined as the its posterior probability. Another popular parametric method is based on the mixture model that models the data with a mixture of parametric statistical distributions. For example, Ge et al.~\cite{ge2010top} modeled the distribution of travel distance between two locations with multiple Gaussian distributions, where each Gaussian distribution describes the distance distribution of one path. Compared with parametric methods, nonparametric methods make fewer assumptions about the data distribution. Kernel function based model\cite{desforges1998applications} is a nonparametric method that estimates probability density using kernel functions, which make no prior assumptions about the distribution. This method is used in \cite{chiang2017btci, chen2015sensing} to model the distribution of traffic speed data and sharing bike renting data. One-class SVM, as a classification based anomaly detection algorithm, learns the region that contains normal points in feature space using kernel function. The instances outside the learned region are identified as anomalies, and the distance between the instance and the region boundary can be regarded as the anomaly degree. In~\cite{zhang2018detecting}, Zhang \emph{et al.} used one-class SVM to detect anomalous events.

\subsubsection{Urban dynamics pattern based}
While urban anomalies are of various types and have complex effects, normal urban activities usually follow some regular patterns. Based on this assumption, some works take an indirect approach of modeling the normal urban dynamics patterns, and then consider the data instances that can not be well described or represented by normal patterns indicating anomalous events. Urban dynamics pattern mining has been widely studied~\cite{zheng2011urban,zhao2016urban,castro2013taxi}. In the context of urban anomaly detection, there are two major approaches for normal urban dynamics modeling, \emph{i.e.},  statistical models and tensor factorization.
\paragraph{Statistical Model}

\begin{figure}[t]
\centering
\includegraphics[width=0.35\textwidth]{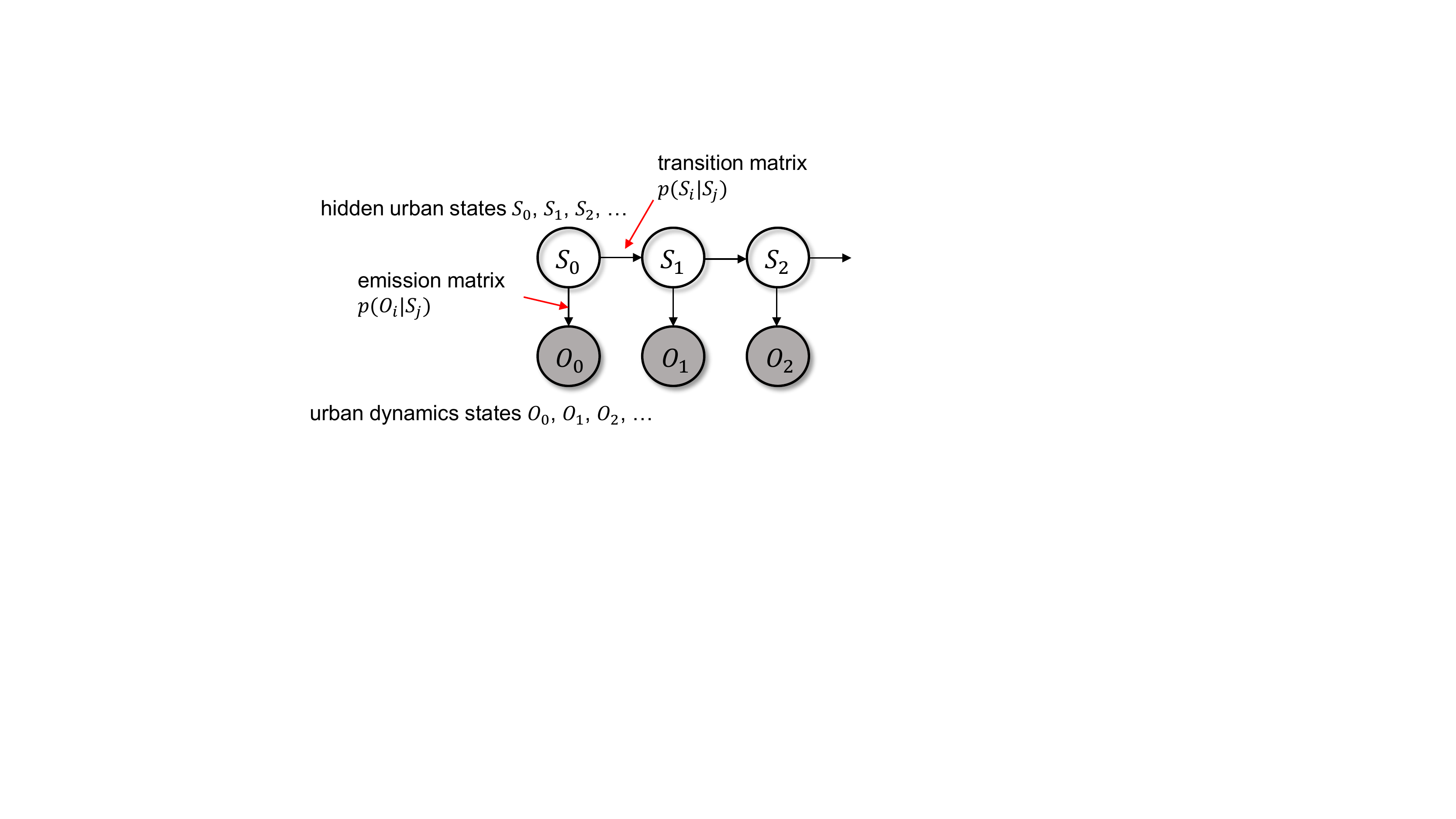}
\caption{The HMM model for urban dynamics transition.}
\label{fig:hmm}
\end{figure}

A common statistical method is to model the normal urban dynamics such as traffic volume~\cite{zhang2015city} and people flow~\cite{ceapa2012avoiding} with a Gaussian distribution, then the probability that anomalous events happen under given an observation is calculated by statistical hypothesis testing methods. Some other works consider there are limited underlying states behind complex urban dynamics and urban anomalies will cause abnormal state transitions. Hidden markov model (HMM) is used to model the urban state transitions. As shown in Fig.~\ref{fig:hmm}, an HMM contains five basic concepts, \emph{i.e.}, observation sequence, state sequence, initial state, transition probability matrix, and emission probability matrix. The observations and states of an HMM are discrete and limited. The hidden state only depends on the previous state, and the observation only depends on the current state. The transition probability matrix is to describe the probability that a state transfer to another state. In the case of urban state modeling, observation sequence corresponds to observed urban dynamics status and state sequence corresponds to hidden urban states. Yang \emph{et al.}~\cite{yang2011anomaly} defined an observation as a vector that represents the number of people and adopted the K-means algorithm to group observation vectors into $k$ clusters as the observed status. Then the initial state and transition probability matrix were estimated based on historical data. To build the emission probability matrix, Yang \emph{et al.} used the Gaussian Mixture Model (GMM) to model the distribution of observation vectors contained in each cluster. After building the model, the probability of a new observation sequence could be calculated. If the probability is under a threshold, an anomalous event is considered happening. Witayangkurn \emph{et al.}~\cite{witayangkurn2013anomalous} proposed a similar algorithm, while they first clustered the observation vectors to reduce the number of observations and defined an anomaly score for each observation instead of an observation sequence at the anomaly detection step.

Likelihood Ratio Test (LRT) is another statistical technique widely applied in spatiotemporal anomaly detection. LRT is originally used to compare models in terms of their statistics. Given a dataset $X$, a model with parameter $\theta \in \Theta_0$, an alternate model with parameter $\theta \in \Theta-\Theta_0$, the likelihood ratio is then defined as,
\begin{equation}
    \lambda = \frac{sup_{\{\theta \in \Theta_0\}}L(\theta|X)}{sup_{\{\theta \in \Theta\}}L(\theta|X)},
\end{equation}
where $\Theta$ is the whole parameter space and $\Theta_0$ is the restricted parameter space. This ratio can be computed by the maximum likelihood estimate (MLE). It is also proved that the asymptotic distribution of $\Lambda = -2\log\lambda$ is a chi-square distribution $\chi^2(\Lambda, p-q)$\cite{wilks1938large}, where $p$ and $q$ are the number of free parameters of the null model and the alternative model respectively. For the anomaly detection, the null model corresponds to the hypothesis that there is no anomaly while the alternate model is corresponding to the opposite case. First, a probability density function will be chosen. Then the value $\Lambda$ can be calculated by MLE. An anomaly is detected with the confidence $\alpha$ when $\Lambda > c$. $c$ is the threshold where the area under chi-square distribution density function is smaller than $\alpha$. Wu \emph{et al.}~\cite{wu2009lrt} proposed a framework to detect the spatial anomaly. The authors first partitioned the spatial area into $n \times n$ grids and checked anomalies for each cell. They made two competing hypotheses on whether the process generating data in a cell is substantially different from the process generating the data outside that cell. Then based on hypotheses, two models with different parameters are proposed. The parameters of the null model are forced to be identical for every cell while the parameters of the alternate model are customized for each cell. Thus, the LRT can be applied to determine which model fits the dataset better. In~\cite{pang2011mining} Pang \emph{et al.} extended the LRT framework in~\cite{wu2009lrt} to discover traffic anomaly and both persistent and emerging outliers can be detected in their work. Khezerlou \emph{et al.}~\cite{khezerlou2017traffic} used the traffic flow to detect gathering events. They represented the traffic flow in an urban area as a directed graph and proposed a definition of the edge anomaly degree based on the likelihood ratio. Finally, they detected gathering events using the anomaly degree of in-edges and out-edges. In~\cite{zheng2015detecting}, LRT is also used to calculate the anomaly degree for regions.

\paragraph{Tensor Factorization}


\begin{figure}[t]
\centering
\includegraphics[width=0.5\textwidth]{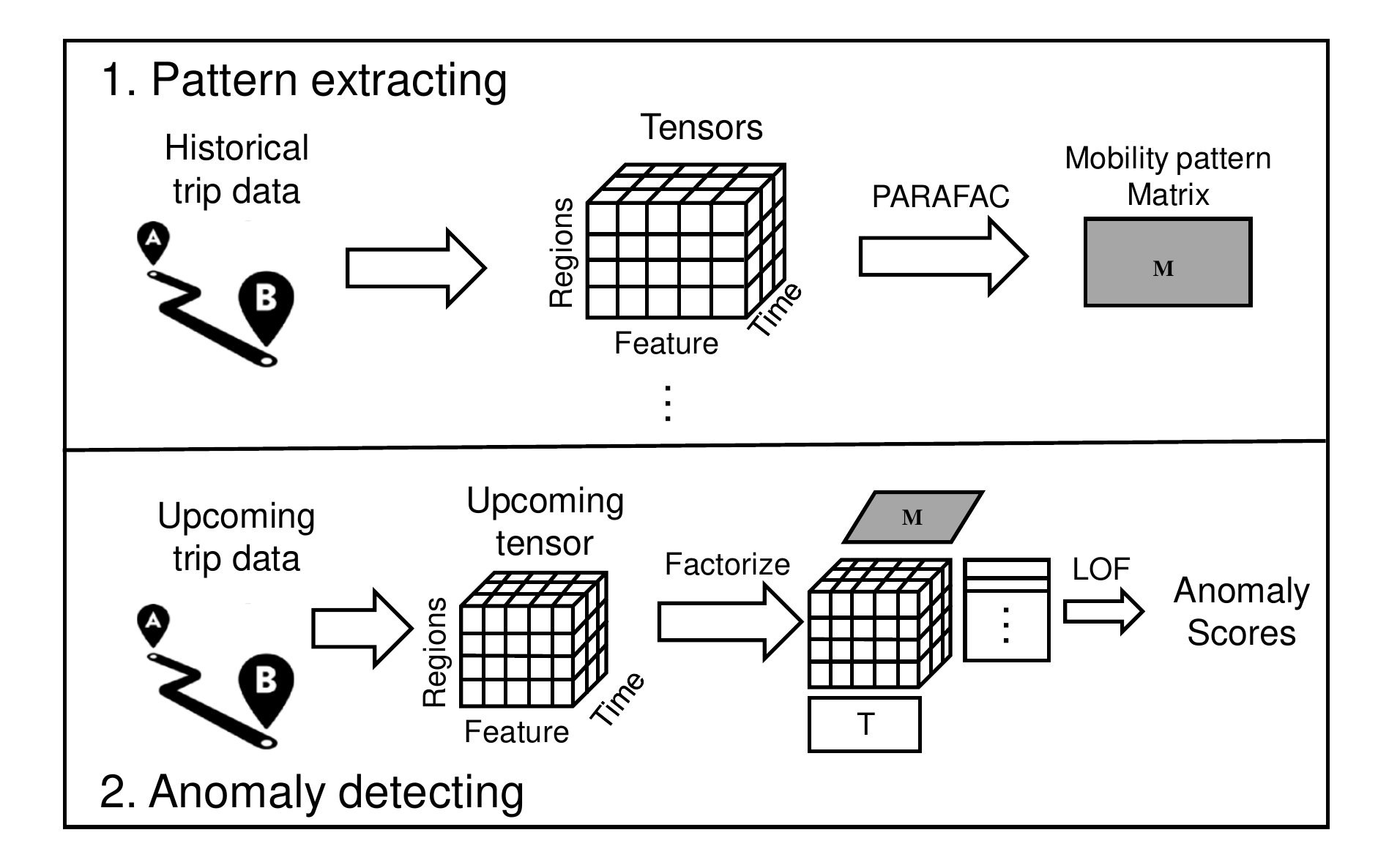}
\caption{The framework of tensor decomposition based method  proposed by Lin \emph{et al.}~\cite{lin2018anomaly}.}
\label{fig:mbexample}
\end{figure}

Some works consider urban dynamics as combination of a certain number of basic urban dynamics patterns that correspond to basic urban human activities such as working, eating and recreating. These patterns can be learned by apply restricted tensor factorization techniques on tensor represented urban data~\cite{fan2014cityspectrum,wang2019understanding,sun2016understanding}. In Fig.~\ref{fig:mbexample} we show a typical example~\cite{lin2018anomaly} of tensor factorization based model. The upper part of Fig.~\ref{fig:mbexample} shows the composition of a region-feature-time tensor, where the feature dimension is the the traffic flow to and from other regions. The lower part of Fig.~\ref{fig:mbexample} shows the detection steps. They first adopted the non-negative CP decomposition~\cite{kolda2009tensor} method to decompose the tensor into a three-factor matrix, which respectively represents the mobility pattern, the temporal and spatial distribution of patterns.  Based on the assumption that urban dynamics in different locations and periods share the same basic mobility patterns, they then decomposed the upcoming tensor with the mobility pattern matrix fixed. In the last, abnormal events were identified in the regions that show abnormal distributions of mobility patterns. In~\cite{chen2017fine}, Chen \emph{et al.} proposed to incorporate social semantic information by cofactorizing a mobility matrix and a social activity check-in tensor together.

\paragraph{Deep neural network}
Deep neural networks have achieved great success on pattern recognition from massive high-dimensional data. Some recent works applied deep neural network models on learning urban dynamic patterns from urban big data. Zhang \emph{et al.}~\cite{zhang2019decomposition} proposed to decompose the urban dynamics into normal and abnormal components, where the former can be learned via a neural network. Trinth \emph{et al.}~\cite{trinh2019urban} adopted Long Short-Term Memory(LSTM) recurrent neural network to model the pattern of urban mobile traffic time series. The effective learning of a deep neural network usually needs sufficient labeled data, which are not accessible in the case of urban anomaly detection. To address this problem, \cite{zhang2019decomposition} proposed to restrict the variance of neural network outputs based on the assumption that normal urban dynamics are stable given the spatiotemporal context. \cite{trinh2019urban} augmented the training data by resampling from the dataset.

\subsubsection{Video anomaly detection}

Surveillance cameras on roads are used to monitor abnormal moving behavior from pedestrian or vehicle flows. However, understanding video data is challenging due to its high dimensionality. Various computer vision methods are designed to capture features from videos for abnormal events detection, ranging from single hand-crafted feature and combined features to representations learned by deep learning models.

In~\cite{Zhang2009Learning, Piciarelli2008Trajectory}, object trajectories in videos were used to describe the mobility of objects. By comparing with normal trajectory patterns, the anomaly events are detected. However, when the object is occluded, or video scenes are crowded, this method would be unable to handle the problem. Therefore, many methods to extract mobility patterns are proposed to overcome this limitation. Benezeth \emph{et al.} ~\cite{Benezeth2009Abnormal} used the histogram of the pixel change; Kim \emph{et al.}~\cite{Kim2009Observe} and Mehran \emph{et al.}~\cite{Mehran2009Abnormal} employed the optical flow to measure dynamic patterns of objects. However, these approaches mainly emphasize dynamics but neglect anomalies of object appearance~\cite{Li2013Anomaly}. For better performances, many features need to be considered together. For example, Li \emph{et al.}~\cite{Li2013Anomaly} and Mahadevan \emph{et al.}~\cite{Mahadevan2010Anomaly} used the Mixture of Dynamic Textures (MDT) models to detect the spatial abnormality as well as the temporal abnormality. Zhu \emph{et al.}~\cite{Zhu2013Context} used the mobility and context features to model the events jointly. Saligrama \emph{et al.}~\cite{saligrama2012video} calculated the anomalous score by aggregating the appearance and mobility features of its nearest neighbor. Additionally, to capture the high-level feature, like the interactions in videos, and the low-level feature, like the motion feature of each video patch, Sabokrou \emph{et al.}~\cite{Sabokrou2015Real} and Cheng \emph{et al.}~\cite{Cheng2015Video} used a hierarchical structure to represent events and interactions. 

\begin{figure*}[t]
    \centering
    \subfigure[Architecture of the model.]{
        \includegraphics[width=0.6\textwidth]{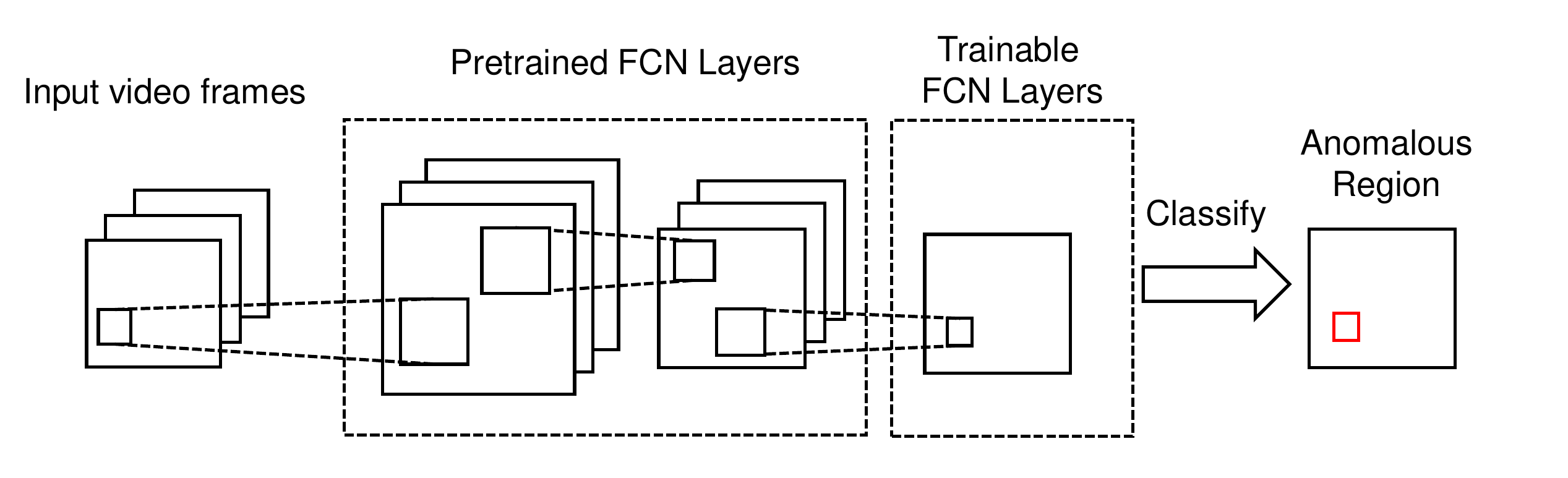}
        \label{fig:cvexample:arch}
    }
    \subfigure[An example of results.]{
        \includegraphics[width=0.17\textwidth]{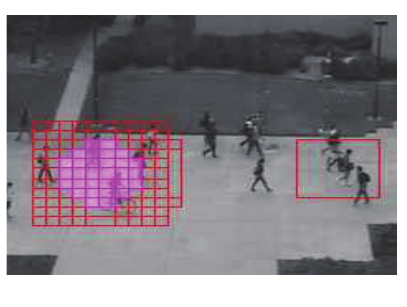}
        \label{fig:cvexample:res}
    }
    \caption{The framework of deep learning based video anomaly detection proposed by Sabokrou \emph{et al.}~\cite{sabokrou2018deep}.}
    \label{fig:cvexample}
\end{figure*}

In recent years, the rapid development of the deep learning methods brings revolution to image and video processing domains including image classification~\cite{Krizhevsky2012ImageNet}, object detection~\cite{Girshick_2014_CVPR} and activity recognition~\cite{Simonyan2014Two}. A lot of modern deep architectures are proposed to replace the hand-crafted features to model activity patterns~\cite{Ravanbakhsh2016Plug}.
Fig.~\ref{fig:cvexample} shows an example of Convolutional Neural Network(CNN) based model~\cite{sabokrou2018deep}. Fig.~\ref{fig:cvexample:arch} shows the architecture of the model, which is a Binary Fully Convolutional Network (BFCN) consists of two pre-trained convolution layers and one output layer. The network is used to capture region features in video frames and the features are further feed into a Gaussian classifier to identify anomalies. An example of detected anomalies is shown in Fig.~\ref{fig:cvexample:res}, in which some pedestrians walking in the opposite direction to other people.
In~\cite{Xu2015Learning, Xu2017Detecting}, appearance and motion features were learned separately by using stacked denoising autoencoders. 
In~\cite{Hasan2016Learning}, a fully connected autoencoder was used on optical-flows, and a fully convolutional feed-forward autoencoder was added to learn both local features and classifiers to capture the temporal regularity of video sequences. 
However, as the size of existing datasets with ground truth abnormality samples are small, the Deep Neural Network based methods have the problem that their networks are relatively shallow to prevent overfitting. 
Also, Generative Adversarial Networks (GANs) \cite{NIPS2014_5423} are introduced for the task of anomaly event detection. In \cite{Ravanbakhsh2017Abnormal}, Ravanbakhsh \emph{et al.} trained GANs with normal data only. Thus, GANs are unable to generate abnormal events. By computing the difference between the real video clips with the representations of appearance and motion reconstructed by the GANs, abnormal areas can be detected.

\subsection{Prediction}
\label{subsec:prediction}
The prediction of urban anomalous events is also a challenging task, and there are many researchers making efforts to forecast urban anomalies. 
A lot of existing works collect real-time urban dynamics to infer whether an anomaly will happen in the near future, especially in the case of traffic anomaly prediction. Besides, some works focus on environment anomalies. Instead of predicting the exact time of anomalous events, these works evaluate whether there is a risk of a certain type of anomalies based on observed features. 
Classification methods are usually used in these two kinds of works, which is summarized in section \ref{sec:classi}. 
Additionally, some other works predict overall distributions of different anomalies by exploring rules from recorded events. Predicting methods such as time series forecasting and deep neural networks are adopted in these works and summarized as regression methods in this section \ref{sec:regress}.

\subsubsection{Classification methods}
\label{sec:classi}

In studies of environment anomaly and traffic anomaly prediction, 
classification methods are usually adopted. The crucial step of making a successful classification is to construct and select appropriate features. In the case of environment anomaly prediction, features are usually constructed from environment information. Madaio \emph{et al.}~\cite{madaio2016firebird} proposed a framework to evaluate the fire risk in Atlanta. They used around 20,000 commercial properties such as fire permits, criminal record, and liquor license to construct features. In \cite{singh2018dynamic}, Singh Walia \emph{et al.} considered the difference between areas in urban functions and selected commercial and residential features respectively. On the other hand, some works tried to predict the water system pollution in cities, and the residents' family information, health condition, and land information were used as features~\cite{chojnacki2017data, abernethy2018activeremediation, potash2015predictive}. After selecting features, classic classification methods can be directly used in these works, including Logistic regression(LR)~\cite{hosmer2013applied}, Support Vector Machine (SVM)~\cite{cortes1995support}, Random Forest~\cite{breiman2001random} and gradient tree boosting~\cite{friedman2001greedy}. In the case of traffic anomaly detection, the road conditions observed by loop detectors and weather information are usually utilized as features. Abdel \emph{et al.}~\cite{abdel2006calibrating} combined weather information and the statistic features of road conditions like the mean speed of vehicles. Xu \emph{et al.}~\cite{xu2013genetic} first adopted Random Forest to select important features~\cite{harb2009exploring} and made a classifier using a Genetic Programming Model~\cite{koza1994genetic}.  In~\cite{xu2013predicting}, Xu \emph{et al.} employed a sequential logistic regression model to predict the severity of traffic accidents. Moreover, Yu \emph{et al.}~\cite{yu2014utilizing} explored the critical factors of different car crash types and then utilized the hierarchical logistic regression model~\cite{wong1985hierarchical} to predict traffic crashes. Deep learning models are also adopted to predict the happening of different kinds of urban anomalies. In~\cite{sun2017dxnat}, Sun \emph{et al.} first mapped traffic data to images and then applied CNN to predict the happening of congestions. Based on the records of anomalous events such as crime and illegal parking, some works focused on predicting the happening of different categories of anomaly at different regions in a city. In~\cite{huang2016crowdsourcing}, Huang \emph{et al.} made the prediction by exploring both the spatial dependency of anomaly occurrence among regions and the historical anomaly distribution of an individual region. In~\cite{huang2018deepcrime}, the spatiotemporal and categorical signals are all embedded into hidden representations and the prediction is made by an attentive hierarchical recurrent network. In~\cite{huang2019mist}, Huang \emph{et al.} further integrated a multi-modal fusion module
and a hierarchical recurrent network to model the spatiotemporal and cross-categorical correlations among crime records data.

\subsubsection{Regression methods}
\label{sec:regress}
In stead of predicting the happening of a single anomalous event, some works adopt regression methods to predict the number of anomalies happening in an urban region in a future time slot. Wu \emph{et al.}~\cite{wu2017uapd} represented urban anomaly records with a tensor. They then factorized the tensor into three factors, \emph{i.e.}, region, category and time matrices and assumed region and category matrices were constant with time. By applying the vector autoregression~\cite{hamilton1994time} algorithm, the next column of the time matrix can be estimated. In the last, by reconstructing the tensor with updated time matrix, the number of anomalies in different regions in the next time step can be predicted. In~\cite{wang2016crime}, Wang \emph{et al.} exploited the Linear Regression and Negative Binomial Regression~\cite{gardner1995regression} model to predict the crime rate of neighborhoods in a city based on both demographic and geographic features. Additionally, deep learning models are also introduced for predicting the number of urban anomalies. By dividing the urban area into grid regions and representing urban dynamics in all regions as a matrix or tensor, the deep learning models that make great achievements in image processing domain can be migrated to deal with urban dynamics. Ren \emph{et al.}~\cite{ren2018deep} proposed LSTM network~\cite{hochreiter1997long} based neural network to predict the risk of traffic accidents in regions based on historical records. Similarly, Yuan \emph{et al.}~\cite{yuan2018hetero} utilized the Convolutional LSTM network~\cite{xingjian2015convolutional} to predict the number of traffic accidents in different regions, which can capture both spatial and temporal domain correlations. In~\cite{chen2016learning}, Chen \emph{et al.} further combined GPS trajectory data and traffic accident data to learn representations of human mobility with a stacked Denoise Autoencoder~\cite{vincent2010stacked} for traffic accident prediction.
\section{Open Challenges and Problems}
\label{sec:chall}
Urban big data-based techniques are the future and promising direction of urban anomaly analysis. 
A great number of works have been done in recent years and obtained a lot of achievements. However, there are still several open problems that have not been well addressed, such as the precise prediction and underlying problem diagnosis. The causes of the problems are essentially the difficulties brought by the complexity of urban big data. To find potential solutions, it is important to investigate the characteristics of urban big data. Therefore, in this section we will first discuss the challenges posed by urban big data and then present consequential open problems of urban anomaly analysis.

\subsection{Data Challenges}
\subsubsection{Data Variety}
Urban data come from different sources are in a verity of forms as introduced in Section \ref{sec:datatype}. To comprehensively discover and understand urban anomalous events, different types of urban data from multiple views need to be combined together. For example, to determine whether the number of people enter a place of interest is overloaded, the video of pedestrian flows and vehicle flows detected on roads are both needed. However, the data volume, processing efficiency and analysis techniques of different types of data can significantly vary from each other, which makes it hard to make use of data in multiple forms together. Moreover, different data sources usually suffer from the spatial and temporal misalignment problem due to different sample rates and sensing areas. To get an accurate snapshot of a specific location and time point, data produced by different sources need to be aligned spatially and temporally, which brings extra difficulty for precise urban anomaly detection. Multi-modal fusion is a potential solution to the data variety problem, which has been applied in urban applications such as traffic prediction~\cite{adetiloye2019multimodal,yao2018deep}. However, the existing works are limited on offline processing and spatiotemporal aligned data. The real-time fusion and spatiotemporal alignment are remained as unsolved problems.

\subsubsection{Data Imbalance}
Although normal urban events happen everyday and everywhere, the anomalous events rarely occur and are usually not recorded. Hence, most of urban data are produced by normal events, and merely a tiny part of human daily activities happening in urban areas cause anomalous events. This extreme imbalance of dataset brings problems from two aspects. First, due to the complex types of anomalies, it is hard to capture patterns of such events and evaluate their influence on urban data. Second, with few recorded anomalous events, it is also hard to evaluate a practical detecting system. Many researchers used typical events such as important festivals and concerts as targets to test the hit rate of their methods. However, since these events are just a subset of the anomalies in real world, it is still far away from a systematical evaluation. Moreover, some researchers use synthetic datasets to evaluate their algorithms. However, the mechanism urban data are produced in real world is extremely complex. It is nearly impossible to simulate the effect of real-world anomalies by simple rules. To remedy the lack of urban anomaly data, cross-city data transfer is a feasible direction. While different cities usually have completely different physical environments, the impact diffusion process of urban anomalies in cyberspace share similar patterns. Transferring data from different cities can greatly enrich the anomalous records and help to discover and model the common patterns. There are already several works trying to combine data from multiple cities to address urban problems such as crowd flow prediction~\cite{wang2018cross} and ridesharing detection~\cite{wang2019ridesharing}. However, combining cross-city data for urban anomaly analysis is remaining as a blank area.

\subsubsection{Data Dependency}
Most data mining and machine learning algorithms assume data points are sampled from independent identical distributions. However, this assumption does not hold for urban big data. Urban data points are usually associated with timestamps and location tags, which bring complex spatial and temporal dependency among them. The spatiotemporal dependency among urban data points lead to difficulties for urban anomaly analysis in two aspects. First, it makes the distribution of normal urban data vary over time and locations. For example, the normal traffic volume on peak hours is usually extremely high if changing the time to midnight. Second, an anomalous event can cause jointly abnormal changes of urban data in different time, locations and sources through the spatiotemporal dependency, making it difficult to diagnose the root cause. For example, the blocking of one road may cause traffic overload on other roads. It is hard to trace back to the underlying problem since the causality among the changes are implicit. The probabilistic graphical model is a natural tool to deal with the dependency among random variables, but it cannot be applied on urban data due to the high dimensionality and unstructured forms. Recent advances of deep casual inference methods~\cite{Scholkopf2019CausalityFM} strive to learn variables and their dependency relations automatically from data, which have been applied on high-dimension unstructured data such as images~\cite{lopez2017discovering}. Accordingly, developing deep spatiotemporal casual inference models is a potential chance to address the dependency challenge of urban big data. 

\subsection{Open Problems}
While the unique and complex qualities of urban big data pose the fundamental and theoretic challenges of data-driven urban anomaly analysis, there are also several open problems that have not been addressed to build practical urban anomaly detection and prediction systems. 
\subsubsection{Reliable Detection}
 Reliability is an essential requirement for urban anomaly detection. False positive and false negative reports can both lead to wrong decisions and cause severe consequences. To make sure the reliability of an urban anomaly detection system, there are two major difficulties. First, the impacts of urban anomalies usually compose of components that are reflected by different types of urban data.
 For example, the audiences attending a concert can choose different types of transportation such as subway, taxi or shared bicycles. Merely depending on part of these data sources may lead to underestimation of the size of concert or even failure to report the event.
 Therefore, to avoid false negative reports, multiple data sources must be effectively combined. Second, while detecting urban anomalies mainly rely on identifying outliers from urban data, the urban data outliers do not necessarily imply urban anomalies. Practical problems such as physical failures of sensors or Internet fake information can also produce abnormal urban data. To avoid false alerts, effective mechanisms must be developed to distinguish these intrinsic outliers of urban data from real urban anomalies. 
 
 \subsubsection{Precise Prediction}
There are some works making efforts on evaluating the risk of accidents and predicting the accumulated number of anomalous events as discussed in section~\ref{subsec:prediction}. However, the precise prediction of a single event is still a blank area. In practice, it is of great demand to the predict the precise time and location of anomalous events in order to take proper actions, especially in the case of crimes or fire risk. However, there are three main difficulties to achieve this goal. First, urban data are full of noises due to complex urban environment. The slight signs of anomalous events in their early stage can be easily drowned in noises. Moreover, the precursors of anomalies sometimes show up in different urban data sources. For example, a protest event that has been discussed a lot on social media platform may cause a traffic blocking event. To make precise prediction, information across multiple datasets need to be linked. At last, after the early signs of urban anomalies been detected, it is still a challenge to infer the precise location and time due to the complexity of the spatial and temporal diffusion process of urban events.

\subsubsection{Problem Diagnosis}
The final goal of urban anomaly analysis is to avoid the happening of anomalies or control and reduce the effect of such events. The gap between existing researches and the final goal is the diagnosis of the underlying problems. While urban anomaly detection and prediction helps us be aware of the happening of anomalous events, diagnosing the root causes help to decide what kind of actions should be taken. For example, an unexpected crowd anomaly can be caused by many reasons, such as a concert, a celebration parade or a terrorist attack. These three underlying events are on different emergency levels and should be handled in different ways. However, finding the underlying problems is difficult because the causes of urban anomalies and the effect of urban anomalies can be reflected by different kinds of urban data and have significant spatial and temporal misalignment. To trace the root causes of urban anomalies, the underlying causality relations needs to be inferred from multi-modal data across time, regions and domains.

\section{Conclusion} \label{sec:Conclusion}
The explosion of urban big data has brought new opportunities to solve traditional urban problems. In this paper, we discussed new rising research areas of big urban data based urban anomaly analytics. To give a comprehensive introduction and literature review of this topic, we studied a considerable number of relevant works in recent years and answered three questions: what kinds of urban data are commonly used in urban anomaly detection and how to represent them? What kinds of anomalous events can be detected or predicted? And what are the general detection and prediction methods? In the last, we summarized the shortcomings of current researches and discussed open problems in this field.

\section*{Acknowledgment}
This research has been supported in part by the project 16214817 from the Research Grants Council of Hong Kong, project FP805 from HKUST, the 5GEAR project and the FIT project from the Academy of Finland, the National Key Research and Development Program of China under grant 2018YFB1800804, the National Nature Science Foundation of China under grant U1836219, 61971267, 61972223, 61861136003, Beijing Natural Science Foundation under grant L182038, Beijing National Research Center for Information Science and Technology under grant 20031887521, and research fund of Tsinghua University-Tencent Joint Laboratory for Internet Innovation Technology. 

\bibliographystyle{unsrt}
\bibliography{bibliography} 

\begin{IEEEbiography}[{\includegraphics[width=1in,height=1.25in,clip,keepaspectratio]{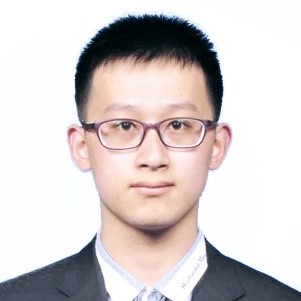}}]{Mingyang Zhang}
 is currently pursuing the Ph.D. degree with the department of Computer Science and Engineering, Hong Kong University of Science and Technology, within the System and Media Laboratory (SymLab). He received the B.S. degrees in electronic engineering from Tsinghua University, Beijing, China, in 2018. His research interests include spatiotemporal data mining, urban computing.
\end{IEEEbiography}


\begin{IEEEbiography}[{\includegraphics[width=1in,height=1.25in,clip,keepaspectratio]{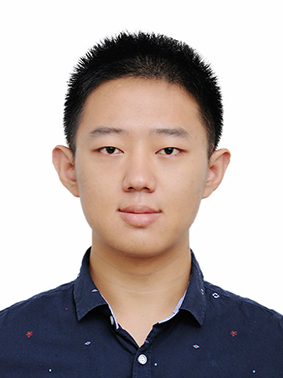}}]{Tong Li} received the B.S.\ degree and M.S. degree in communication engineering from Hunan University, China, in 2014 and 2017.
At present, he is a dual Ph.D. student at the Hong Kong University of Science and Technology and the University of Helsinki. \ His research interests include data mining and machine learning, especially with applications to mobile big data and urban computing. He is an IEEE student member.\
\end{IEEEbiography}


\begin{IEEEbiography}[{\includegraphics[width=1in,height=1.25in,clip,keepaspectratio]{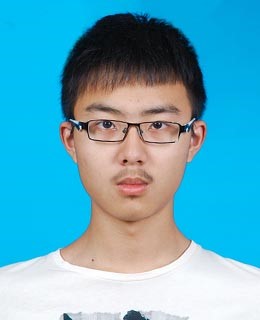}}]{Yue Yu}
is a Ph.D student in School of Computational Science and Engineering, Georgia Institute of Technology, Atlanta, USA. His work mainly focuses on spatio-temporal data mining and machine learning.
\end{IEEEbiography}


\begin{IEEEbiography}[{\includegraphics[width=1in,height=1.25in,clip,keepaspectratio]{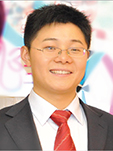}}]{Yong Li}(M'09-SM'16) received the B.S. degree in electronics and information engineering from Huazhong University of Science and Technology, Wuhan, China, in 2007 and the Ph.D. degree in electronic engineering from Tsinghua University, Beijing, China, in 2012. He is currently a Faculty Member of the Department of Electronic Engineering, Tsinghua University.

Dr. Li has served as General Chair, TPC Chair, SPC/TPC Member for several international workshops and conferences, and he is on the editorial board of two IEEE journals. His papers have total citations more than 6900. Among them, ten are ESI Highly Cited Papers in Computer Science, and four receive conference Best Paper (run-up) Awards. He received IEEE 2016 ComSoc Asia-Pacific Outstanding Young Researchers, Young Talent Program of China Association for Science and Technology, and the National Youth Talent Support Program.
\end{IEEEbiography}


\begin{IEEEbiography}[{\includegraphics[width=1in,height=1.25in,clip,keepaspectratio]{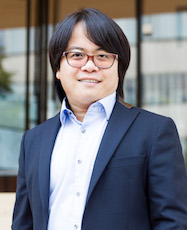}}]{Pan Hui}
(SM'14-F'18) received his Ph.D. degree from
the Computer Laboratory at University of Cambridge, and both his Bachelor and MPhil degrees from the University of Hong Kong.

He is the Nokia Chair Professor in Data Science and Professor of Computer Science at the University of Helsinki. He is also the director of the HKUST-DT Systems and Media Lab at the Hong Kong University of Science and Technology. He was a senior research scientist and then a Distinguished Scientist for Telekom Innovation Laboratories (T-labs) Germany and an adjunct Professor of social computing and networking at Aalto University.  His industrial profile also includes his research at Intel Research Cambridge and Thomson Research Paris. He has published more than 300 research papers and with over 17,500 citations. He has 30 granted and filed European and US patents in the areas of augmented reality, data science, and mobile computing. He  has been serving on the organising and technical program committee of numerous top international conferences including ACM SIGCOMM, MobiSys, IEEE Infocom, ICNP, SECON, IJCAI, AAAI, ICWSM and WWW. He is an associate editor for the leading journals IEEE Transactions on Mobile Computing and IEEE Transactions on Cloud Computing. He is an IEEE Fellow, an ACM Distinguished Scientist, and a member of the Academia Europaea.\
\end{IEEEbiography}


\begin{IEEEbiography}[{\includegraphics[width=1in,height=1.25in,clip,keepaspectratio]{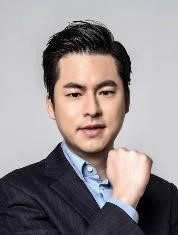}}]{Yu Zheng}is the Vice President and Chief Data
Scientist at JD Finance, passionate about using
big data and AI technology to tackle urban chal-
lenges. He is the general manager of the Urban
Computing Business Unit and serves as the direc-
tor of the Urban Computing Lab at JD Group.
Before Joining JD Group, he was a senior research manager at Mi-
crosoft Research. Zheng is also a Chair Professor at Shanghai Jiao
Tong University, an Adjunct Professor at Hong Kong University of
Science and Technology. Zheng currently serves as the Editor-in-
Chief of ACM Transactions on Intelligent Systems and Technology
and has served as chair on over 10 prestigious international confer-
ences, e.g. as the program co-chair of ICDE 2014 (Industrial Track),
CIKM 2017 (Industrial Track), IJCAI 2019 (Industrial Track). In 2013,
he was named one of the Top Innovators under 35 by MIT Technolo-
gy Review (TR35) and featured by Time Magazine for his research
on urban computing. In 2014, he was named one of the Top 40 Busi-
ness Elites under 40 in China by Fortune Magazine. In 2017, Zheng is
honored as an ACM Distinguished Scientist.
\end{IEEEbiography}

\end{document}